\documentclass[aps,prd,notitlepage,amssymb,amsmath,floatfix,nofootinbib,superscriptaddress]{revtex4-1}
\usepackage{amsmath,graphicx,epsfig,amssymb,dsfont,mathtools}
\usepackage[usenames]{color}

\allowdisplaybreaks[4]

\usepackage{subcaption}

\begin{document}
\title{Maximal flavor violating $U(1)_{L_{\mu}-L_{\tau}}$ model with singly-charged scalar}
\author{Fei Huang}
\affiliation{School of Physics and Technology, University of Jinan, 250022, Jinan, China,}
\author{Jin Sun}
\email{corresponding author: sunjin0810@ibs.re.kr}
\affiliation{Particle Theory and Cosmology Group, Center for Theoretical Physics of the Universe, Institute for Basic Science (IBS), Daejeon 34126, Korea}

\preprint{CTPU-PTC-24-30}

\begin{abstract}
The $U(1)_{L_\mu-L_\tau}$ $Z'$ model has emerged as a promising candidate to address the longstanding muon $(g-2)_\mu$ anomaly.
Flavor-conserving $Z'$ interactions are subject to stringent constraints from the neutrino trident process and NA64$\mu$ experiments, which limit the $Z'$ mass to $m_{Z'}<40$ MeV.
To circumvent these constraints, flavor-conserving $Z'$ interactions  can be converted into maximal flavor-violating interactions through a discrete exchange symmetry.
Maximal flavor-violating $Z'$ interactions contribute to $\tau\to\mu\nu\bar\nu$ via the neutral current, yet the parameter space for this process conflicts with the $(g-2)_\mu$ allowed regions.
To resolve this conflict, we propose introducing a singly-charged scalar that mediates via charged current interactions. This scalar is anticipated to produce a negative contribution to the lepton $(g-2)_l$ while concurrently inducing the tau decay $\tau\to\mu\nu\bar\nu$.
Three distinct scenarios arise from the introduction of singly-charged scalars: the $\mathcal{O}_{LL}$ operator, driven by weak singlet and triplet, and the $\mathcal{O}_{LR}$ operator, driven by weak doublet.
Our analysis of the phenomenology in these three cases reveals that the tension between the muon $(g-2)_\mu$ anomaly and $\tau\to\mu\nu\bar\nu$ for large $Z'$ masses can be effectively alleviated only by the singly-charged scalars from the weak triplet, whereas the singlet and doublet scenarios fall short.
Furthermore, the singly-charged scalars from the weak triplet offer an additional explanation for the electron $(g-2)_e$ anomaly.
Our findings indicate that weak triplets could play a  crucial  role in $Z'$ models, potentially providing valuable insights for future research into $U(1)$ frameworks.
\end{abstract}
\maketitle

\section{Introduction}
\label{sec:intro}
The Standard Model (SM) of particle physics has proven to be remarkably successful in accurately describing elementary particles and their interactions.
However, in recent years, significant discrepancies have emerged between Standard Model predictions and observed flavor observables~\cite{Crivellin:2023zui}. 
Among the most notable of these discrepancies is the long-standing anomaly in the muon anomalous magnetic moment, $a_\mu \equiv (g-2)_\mu/2$~\cite{Muong-2:2021ojo,Muong-2:2023cdq}. Recently, a new approach known as the “window” observable theory~\cite{Ce:2022kxy,Colangelo:2022vok} has been proposed to reconcile the differences between dispersive and lattice methods for computing the muon $(g-2)_\mu$. 
The persistent deviation in $a_\mu$ is widely regarded as strong evidence for physics beyond the Standard Model (BSM); for a comprehensive review, see Refs~\cite{Jegerlehner:2009ry,Lindner:2016bgg,Athron:2021iuf}. 
A promising candidate for explaining this anomaly is the gauged $U(1)_{L_\mu - L_\tau}$ $Z'$ model~\cite{He:1990pn,Baek:2001kca,Ma:2001md,Altmannshofer:2014pba,Heeck:2016xkh,Altmannshofer:2016oaq,Amaral:2021rzw,Zu:2021odn,Kang:2021jmi}. 
This $Z'$ interaction with muons is also expected to alter the muon neutrino trident (MNT) process $\nu_\mu N \to \nu_i \mu^+\mu^- N$, placing constraints on the $Z'$ mass to be below approximately 300 MeV if it is to resolve the $(g-2)_\mu$ anomaly~\cite{Altmannshofer:2014pba,Cen:2021ryk}.
Recently, the NA64$\mu$ experiment, employing the missing energy-momentum technique with a muon beam, has imposed significantly tighter constraints, limiting the $Z'$ mass to $m_{Z'}<40$ MeV~\cite{NA64:2024klw}. 
These stringent bounds can be circumvented by converting the original flavor-conserving $Z'$ interaction, $(\bar \mu \gamma^\mu \mu - \bar \tau \gamma^\mu \tau + \bar \nu^\mu_L \gamma^\mu \nu^\mu_L - \bar \nu^\tau_L \gamma^\mu \nu^\tau_L)Z'_\mu$, into a fully off-diagonal interaction, $(\bar \mu \gamma^\mu \tau + \bar \tau \gamma^\mu \mu + \bar \nu^\mu_L \gamma^\mu \nu^\tau_L + \bar \nu^\tau_L \gamma^\mu \nu^\mu_L)Z'_\mu$, through discrete symmetry transformations~\cite{Foot:1994vd,Liu:2024gui}. However, in scenarios with a large $Z'$ mass, these flavor-violating interactions are excluded by the stringent bounds on the decay process $\tau\to\mu \nu\bar\nu$. 
Fortunately, the allowed mass range for $Z'$ can be expanded in the context of resolving the muon $g-2$ anomaly by introducing type-II seesaw $SU(2)_L$ triplet scalars, as shown in Refs.~\cite{Cheng:2021okr,Sun:2023ylp}. Specifically, the introduction of a singly-charged scalar from the triplet can mitigate this tension.
However, the accompanying doubly-charged scalar complicates the analysis, as it mediates processes like $\tau \to 3\mu,\;\mu \gamma$ and  is tightly constrained by collider experiments. 
To streamline the analysis, one might focus exclusively on the singly-charged scalar, which has been extensively employed in analyzing the decay $\tau\to \mu\nu\bar\nu $~\cite{Crivellin:2020oup,Crivellin:2020klg, Felkl:2021qdn}. Previous studies have demonstrated that a singly-charged scalar within the triplet scalar framework can substantially extend the viable $Z'$ mass range~\cite{Cheng:2021okr}. 
This raises the question of whether alternative mechanisms exist to achieve similar effects, given that singly-charged scalars can originate from various sources.

The singly-charged scalar can emerge from various extensions of the Standard Model (SM) that incorporate additional Higgs multiplets. 
The most straightforward extension is the Two Higgs Doublet Model (2HDM), which features a physical charged Higgs boson $H_D^\pm$ whose couplings to fermions are proportional to their masses, as discussed in the reviews~\cite{Diaz:2002tp,Branco:2011iw}. 
The type-II seesaw mechanism~\cite{Magg:1980ut,Cheng:1980qt,Schechter:1980gr,Lazarides:1980nt,Mohapatra:1980yp}, which introduces scalar triplets to explain small neutrino masses at tree level, also provides a physical charged scalar and can simultaneously lead to deviations of the electroweak $\rho$ parameter $\left(\rho = \frac{m_W^2}{m_Z^2\cos^2 \theta_W}\right)$ from unity. 
The Georgi-Machacek model~\cite{Georgi:1985nv}, which introduces both complex and real weak triplets, predicts two singly-charged Higgs bosons that transform as a 3-plet and a 5-plet under $SU(2)_C$ while preserving $\rho=1$ at tree level. 
The couplings of $H_3^\pm$ to fermions and $H_5^\pm$ to $W^\mp Z$ are determined by the triplet vacuum expectation value (VEV). Additionally, weak-singlet singly-charged scalars are featured in the Zee~\cite{Zee:1980ai} and Zee-Babu models~\cite{Zee:1985id,Babu:1988ki}, which generate small radiative neutrino masses through one- and two-loop processes, respectively.

Experimental searches for singly-charged Higgs bosons have been extensively carried out at high-energy colliders, exploiting their distinctive electric charge signature. 
The specific strategies for these searches are shaped by the couplings between the charged Higgs and Standard Model (SM) particles. 
Within the 2HDM framework, LEP experiments have ruled out charged Higgs bosons with masses below 80 GeV (Type-II) and 72.5 GeV (Type-I)~\cite{ALEPH:2013htx}. 
At the LHC, searches for charged Higgs bosons in the 2HDM primarily focus on top quark decays when $m_{H^\pm_D}<m_t$, or on associated production with a top quark for $m_{H^\pm_D}>m_t$, where $m_t$ denotes the top quark mass.
Numerous decay channels have been probed for $90\text{GeV}<m_{H^\pm_D}<2$ TeV, but no significant excess has been detected. These channels include $tb$~\cite{ATLAS:2018ntn,ATLAS:2015nkq}, $\tau\nu$~\cite{ATLAS:2018gfm,ATLAS:2014otc}, $cs$~\cite{CMS:2015yvc}, and $cb$~\cite{CMS:2018dzl}. 
In the Georgi-Machacek (GM) model, the singly-charged Higgs has been investigated through vector-boson fusion processes such as $pp \to jj H_5^\pm(\to W^\pm Z)$ for $90\text{GeV}<m_{H_5^\pm}<1$ TeV\cite{ATLAS:2015edr} and resonant $WZ$ production in fully leptonic final states~\cite{ATLAS:2018iui}. 
A global fit of all LHC searches for charged Higgs bosons within the GM model indicates that the triplet's charged member remains viable for masses above 200 GeV~\cite{Chiang:2018cgb}. Additionally, the LHC's sensitivity to weak-singlet singly charged scalars decaying into electrons and muons has been evaluated in Ref.~\cite{Alcaide:2019kdr}. 
Based on these experimental studies, we consider the singly-charged scalar  with hundreds of GeVs in the subsequent analysis.

In this manuscript, we construct an ultraviolet (UV) model featuring a maximal lepton flavor-violating $Z'$ interaction by incorporating three Higgs doublets with distinct $U(1)_{L_\mu-L_\tau}$ charges and enforcing a discrete exchange symmetry. 
The introduction of a singlet scalar allows for an increase in the mass of the gauged boson $Z'$. While the maximal flavor-violating $Z'$ interaction induces the decay $\tau \to \mu \nu \bar{\nu}$, which restricts the model's parameter space, this process is mediated not only by the neutral current interaction of the maximal flavor-changing $Z'$ boson but also by the charged current interaction of the singly-charged Higgs. 
To address the tension between the $(g-2)_\mu$ anomaly and the $\tau \to \mu \nu \bar{\nu}$ decay caused by the maximal flavor-violating $Z'$ interaction, we propose the introduction of  singly-charged scalars.
Generally, the leptonic interactions involving the charged scalar  occur with lepton pairs, irrespective of their chirality. 
However, without including right-handed neutrinos, the interaction is characterized by a single operator in three different scenarios:
\begin{itemize}
  \item  Case I. Weak singlet induced $\mathcal{O}_{LL}$ operator: $\bar{\nu}_{La}e^{c}_{Lb}h^{-}$+h.c.
  \item  Case II. Weak triplet induced $\mathcal{O}_{LL}$ operator: $\bar{\nu}_{La}e^{c}_{Lb}\Delta^{-}$+h.c. 
  \item  Case III. Weak doublet induced  $\mathcal{O}_{LR}$ operator: $\bar{\nu}_{La}e_{Rb}H^{+}$+h.c., 
\end{itemize}
here the superscript $c$  denotes  the charge conjugation transformation, and the subscript L(R) represent the left-handed (right-handed) component.
We investigate the possibility of resolving the $(g-2)_\mu$ anomaly and the $\tau \to \mu \nu \bar{\nu}$ decay issue associated with   the maximal flavor-violating $Z'$ interaction under three different scenarios involving singly-charged scalars.
Our analysis indicates that only Case II effectively resolves the conflict, while Case I worsens the issue, and Case III maintains an approximate situation comparable to the original 
$Z'$ interaction.

The layout of this article is given as follows:
Section II outlines the constraints on the flavor-conserving $U(1)_{L_\mu-L_\tau}$ model, 
then constructs the maximal off-diagonal $Z'$ interactions, and analyzes the viable parameter space for  $(g-2)_\mu$, $\tau\to \mu Z'$ and  $\tau\to \mu \nu\bar\nu$.
Section III explores the possibility of expanding the $Z'$ parameter range by introducing  singly-charged scalars across three different scenarios and further analyzes the associated phenomenology.
 Section IV draws our conclusion.

%

\section{The $U(1)_{L_\mu-L_\tau}$  model with flavor-conserving or -changing $Z'$ interaction}
\label{sec:qqbar}

The gauge group of the model is $SU(3)_C\times SU(2)_L\times U(1)_Y\times U(1)_{L_\mu-L_\tau}$ with the corresponding coupling constants $g_s$, $g$, $g'$ and $\tilde g$, respectively.
The left-handed leptons $L_L = (\nu_L,\;l_L)^T = (L_{L_1},\;L_{L_2}, \;L_{L_3})$,  the right-handed charged leptons, $l_R = (l_{R_1},\;l_{R_2}, \;l_{R_3})$ have $SU(3)_C\times SU(2)_L\times U(1)_Y$ quantum numbers  $(1,2,-1/2)$ and  $(1,1,-1)$, respectively.
Here the subscripts ${1,2,3}$ correspond to  different generations.
The first, second and third generations transform under the gauged $U(1)_{L_\mu-L_\tau}$ as $0,\;1,\;-1$, respectively.
The corresponding charges for all particles in our model are collected in Table.~\ref{U1charge}.

\begin{table}[]
    \centering
    \begin{tabular}{|c|c|c|c|c|c|c|c|c|c|c|c|c|c|c}
    \hline\hline
Guage Group &$L_{1L}$ &$L_{2L}$ &$L_{3L}$ &$e_{1R}$& $e_{2R}$ &$e_{3R}$ &$H_1$ &$H_2$ &$H_3$ & $S$ & $h_1^\pm$ & $h_2^\pm$ & $h_3^\pm$ \\\hline
$SU(3)_C$ &1 &1 &1& 1 &1 &1 &1 & 1 &1 &1 &1 &1 & 1\\
$SU(2)_L$ &2 &2 & 2&1& 1 & 1&2 &2 &2 & 1 & 1 & 1 & 1 \\
$U(1)_Y$ &-1/2 &-1/2 &-1/2 &-1 & -1 & -1 &1/2 &1/2 &1/2 & 0 &1 & 1 &1 \\\hline
$U(1)_{L_\mu-L_\tau}$ &0 &1&-1& 0 &1 &-1&0 & 2&-2 & 1& 0 & 2&-2 \\\hline
    \end{tabular}
    \caption{particle contents of the model and corresponding  charges under the gauged groups $SU(3)_C\times SU(2)_L\times U(1)_Y\times U(1)_{L_\mu-L_\tau}$. Here $h^{\pm}_{i}$( $i=1,2,3$) represent the singly-charged scalars with ($h_{i}^{\pm},\Delta^{\pm}_{i},H^{\pm}_{i}$) for three different scenarios.}. 
    \label{U1charge}
\end{table}

In this section, we first analyze the flavor-conserving $U(1)_{L_\mu-L_\tau}$ $Z'$ interaction and its associated phenomenology, including $(g-2)_\mu$ and neutrino trident process.
We then proceed with the model construction for maximal flavor-changing $Z'$  interactions and explore the viable parameter regions by considering constraints from $\tau\to \mu \nu\bar\nu$ and $\tau\to \mu Z'$ decays.

\subsection{flavor-conserving $Z'$ interaction}

\begin{figure*}[!t]
    \centering
    \begin{subfigure}[b]{0.45\textwidth}
        \centering
        \includegraphics[width=.8\textwidth]{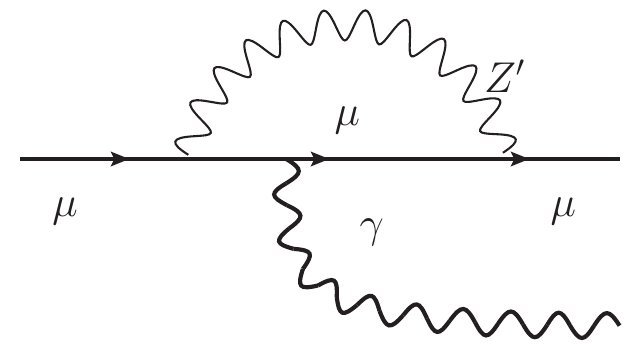}
        \caption{$a_\mu$}
        \label{fig:LFC}
    \end{subfigure}
    \hfill
    \begin{subfigure}[b]{0.45\textwidth}
        \centering
        \includegraphics[width=.8\textwidth]{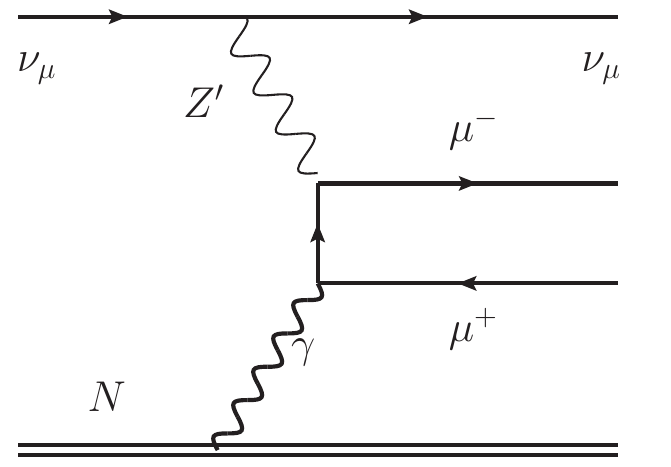}
        \caption{MNT-$Z'$}
        \label{fig:MNT}
    \end{subfigure}
    \vspace{1em}
    \begin{subfigure}[b]{0.45\textwidth}
        \centering
        \includegraphics[width=.8\textwidth]{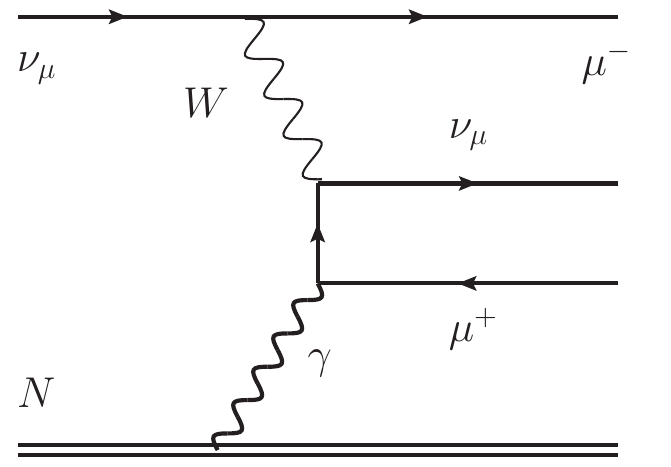}
        \caption{MNT-W}
        \label{fig:MNTW}
    \end{subfigure}
    \hfill
    \begin{subfigure}[b]{0.45\textwidth}
        \centering
        \includegraphics[width=.8\textwidth]{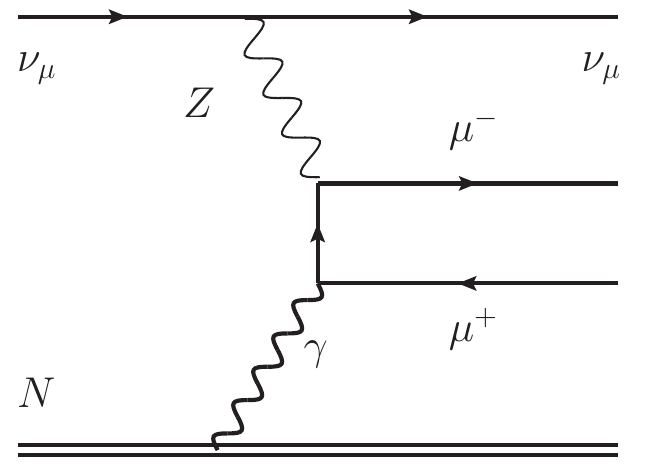}
        \caption{MNT-Z}
        \label{fig:MNTZ}
    \end{subfigure}
    \caption{The Feynman diagrams for $(g-2)_\mu$ and neutrino trident processes with flavor-conserving $Z'$ interaction. The last two diagrams (c,d) represent the SM contribution mediated by W and Z bosons, respectively.}
    \label{fig:LFC-NMT}
\end{figure*}

In the simplest $U(1)_{L_\mu - L_\tau}$ model, 
the $Z^\prime$ gauge boson of the model only interacts with leptons in the  weak interaction basis~\cite{He:1990pn,He:1991qd}
\begin{eqnarray}\label{diagonal}
{\cal L}_{Z'}=- \tilde g (\bar \mu \gamma^\mu \mu - \bar \tau  \gamma^\mu \tau + \bar \nu_\mu \gamma^\mu L \nu_\mu - \bar \nu_\tau \gamma^\mu L \nu_\tau) Z^\prime_\mu \;, \label{zprime-current}
\end{eqnarray}
where  $L(R) = (1 - (+)\gamma_5)/2$.
Note that the interaction in Eq.~\ref{diagonal} is fully flavor-conserving one. 
The vector-type couplings will contribute to muon $(g-2)_\mu$ by exchanging $Z'$ at one loop level~\cite{Leveille:1977rc} as shown in Fig.~\ref{fig:LFC},
\begin{eqnarray}
\Delta a_\mu^{Z^\prime}= \frac{\tilde g^2}{ 8\pi^2} \frac{m^2_\mu}{ m^2_{Z^\prime}} \int^1_0 \frac{2 x^2(1-x) dx}  {1-x + (m^2_\mu/m^2_{Z^\prime}) x^2}\;. \label{leading}
\end{eqnarray}
In the limit $m_{Z^\prime} >> m_\mu$, $\Delta a^{Z^\prime}_\mu = (\tilde g^2/ 12 \pi^2)(m^2_\mu/m^2_{Z^\prime} )$.  
While the experimental measurement of $(g-2)_\mu$ is uncontroversial~\cite{Muong-2:2006rrc,Muong-2:2021ojo,Benayoun:2023dkl},
the corresponding SM theory prediction is subject
to ongoing debate and research, approached through three primary methods.
The first approach is the Muon $(g-2)$ Theory Initiative~\cite{Aoyama:2020ynm}, which employed dispersive techniques to extract the leading-order hadronic vacuum polarization (HVP) contribution from $e^+e^-\to$ hadrons data, resulting in a discrepancy of $5.1\sigma$ between theory and experiment.
In contrast, the BMW Lattice QCD collaboration~\cite{Borsanyi:2020mff} in turn employed a first-principles lattice QCD approach to calculate the HVP contribution to $(g-2)_\mu$, reducing the discrepancy to $1.6\sigma$.
To address the discrepancy between the dispersive and lattice methods, a new approach known as “window” observable theory~\cite{Ce:2022kxy,Colangelo:2022vok} has been proposed. This approach utilizes data-driven techniques that are less susceptible to systematic uncertainties, yielding a discrepancy of $3.8\sigma$.
The corresponding discrepancies between each theory prediction and the experimental measurement read~\cite{Acaroglu:2023cza,Armando:2023zwz}
\begin{align}\label{muong2}
&\Delta a_{\mu}^{\mathrm{exp,dat}}=(2.49\pm0.48)\times10^{-9},\notag\\&\Delta a_{\mu}^{\mathrm{exp,lat}}=(1.05\pm0.62)\times10^{-9},\notag\\&\Delta a_{\mu}^{\mathrm{exp,dat+lat}}=(1.81\pm0.47)\times10^{-9}.
\end{align}
For a summary
of the current status of the  SM prediction for $(g-2)_\mu$, refer to
Refs.~\cite{Stoffer:2023gba,g2}.
In the following, we will use the last one (Window theory), $\Delta a^W_{\mu}=(1.81\pm0.47)\times10^{-9}$,  for our analysis.

If we further consider the kinetic mixing effects, the relevant Lagrangian is 
\begin{eqnarray}\label{mix}
\mathcal{L}=-\frac{1}{4}B_{\mu\nu}B^{\mu\nu}+\frac{\sigma}{2} B_{\mu\nu}Z^{{^{\prime}\mu\nu}}-\frac{1}{4}Z_{\mu\nu}^{\prime}Z^{{^{\prime}\mu\nu}}, 
\end{eqnarray}
here $B$ and $Z'$ stand for  the $U(1)_Y$  and $U(1)_{L_\mu-L_\tau}$ gauged fields, respectively. The parameter $\sigma$ quantifies the  mixing strength with $|\sigma|<0.1$. 
The aforementioned kinetic mixing will introduce a correction to the muon $(g-2)$ at second order
$\sigma^2$~\cite{Cen:2021ryk}
\begin{eqnarray}\label{g2conserving}
   a_\mu^{mix}= \frac{\tilde{g}^2}{12\pi^2}\frac{m_\mu^2}{m_Z^2}\left(\frac{m_Z^2}{m_Z^2-m_{Z^{\prime}}^2}\right)^2\sigma^2\sin^2\theta_W\;.
\end{eqnarray}
For $m_{Z'}>m_{Z}$ and $\sigma<0.1$, the kinetic mixing effect can be ignored.

Besides, the $Z'$
exchange will produce a non-zero contribution to the neutrino trident  process (MNT): the production of a $\mu^+\mu^-$ pair from the scattering of a muon-neutrino with heavy nuclei  $v_\mu +N\to v_\mu +N +\mu^+ \mu^-$. 
The SM contribution includes electroweak W and Z boson as shown in Figs.~(\ref{fig:MNTW}, \ref{fig:MNTZ}),  described by the Lagrangian
\begin{align}
\mathcal{L}_{SM}=-\frac{g^2}{16m_{W}^2}\bar{\mu}\gamma_{\mu}(1+4s_{W}^2-\gamma_{5})\mu\bar{v}_{\mu}\gamma^{\mu}(1-\gamma_{5})v_{\mu}\;.
\end{align}
here  $s_{W}=\sin \theta_{W}$ with  weak Weinberg mixing angle $\theta_W$. 
The Fierz transformation is employed  to combine the W and Z contribution.
In the heavy $Z'$ limit with GeV scale, the $Z'$ model gives the correction to the cross section in Fig.~\ref{fig:MNT} as
\begin{eqnarray}
{\sigma_{Z^\prime}\over \sigma_{SM}} \vert_{trident}= { (1+4s_W^2 + 8 \tilde g^2 m^2_W/g^2 m^2_{Z^\prime})^2 + 1 \over 1 + (1+4 s^2_W)^2}\;, \label{trident-neutrino}
\end{eqnarray}
The current ratio is measured by three  experiments with~\cite{CHARM-II:1990dvf,CCFR:1991lpl,NuTeV:1999wlw},
\begin{eqnarray}\label{tridentexp}
\frac{\sigma(v_\mu \to v_\mu \mu^+\mu^-)^{exp}}{\sigma(v_\mu \to v_\mu \mu^+\mu^-)^{SM}}
=\left\{\begin{array}{clc|}
  1.58\pm 0.57 \;& (\mbox{CHARM-II }) \\
  0.82\pm 0.28\;& (\mbox{CCFR}) \\
  0.72^{+1.73}_{-0.72}\;& (\mbox{NuTeV})
\end{array} \right. \;.
\end{eqnarray}
The above gives the weighted average data $0.95\pm 0.25$. However, the above neutrino trident data excluded the large $Z'$ mass case for resolving the $(g-2)_\mu$ anomaly. In the following we show the relevant details.

 Assuming the difference $\Delta a_\mu$ is due to $\Delta a^{Z^\prime}_\mu$, we have
${\tilde g^2/m^2_{Z'}} = (1.92\pm 0.50)\times 10^{-5} \mbox{GeV}^{-2}$. Inserting this into Eq.~\ref{trident-neutrino}, we obtain
$\sigma_{Z^\prime}/ \sigma_{SM} = 4.05$ which is several times larger than experimentally allowed upper bounds~\cite{CHARM-II:1990dvf, CCFR:1991lpl, NuTeV:1999wlw}.  In this context, $Z^\prime$ contribution to MNT  process should be lowered to a factor 20 to be within the one sigma range of the weighted average $0.95\pm 0.25$ from the above three data.

Therefore, the neutrino trident data constrain the upper limit of $m_{Z'}$ to be less than 300 MeV~\cite{Altmannshofer:2014pba,Cen:2021ryk}.
 And NA64$\mu$ further provides stronger constraints around  $m_{Z'}<40$MeV~\cite{NA64:2024klw}.

Therefore, we attempt to find the ways to evade these stringent constraints.
If there is a mixing $\theta$ between $\mu$ and $\tau$, the MNT process will be suppressed by a factor of $\theta^2$, and  simultaneously the parameter ${\tilde g^2/m^2_{Z'}}$ is suppressed by the factor $\theta^2 m_\tau/m_\mu$. 
This reduction in the parameter can lessen the MNT constraint.
However, with both flavor-diagonal and off-diagonal $Z'$ couplings to $\mu$ and $\tau$, lepton flavor violating (LFV) processes, such as $\tau \rightarrow \mu \gamma$ and $\tau \rightarrow 3 \mu$, will be induced.
These processes impose very stringent constraints, making it impossible for the mixing model to explain the $(g-2)_\mu$ anomaly~\cite{Cen:2021ryk}.
Therefore, if the flavor-conserving interaction described in Eq.~\ref{diagonal} can be converted into a purely flavor-changing interaction, the stringent bounds from neutrino trident and NA64$\mu$ could be avoided while still prohibiting lepton flavor violating (LFV) processes.
Fortunately, such a purely flavor-changing interaction can be realized as proposed some time ago~\cite{Foot:1994vd}.
In the  proceeding subsection, we will present the ultraviolet (UV) realization of $U(1)_{L_\mu-L_\tau}$ with the maximal off-diagonal $Z'$ interaction.

\subsection{flavor-changing $Z'$ interaction}

The off-diagonal $Z'$ interaction can be realized by introducing new scalar particles, as shown in Refs.~\cite{Foot:1994vd,Cheng:2021okr}.
In order to construct the maximal flavor-changing $Z'$ interaction, three Higgs doublets $H_{1,2,3}: (1,2, 1/2)$ ($<H_i> = v_{i}/\sqrt{2}$) with   $U(1)_{L_\mu - L_\tau}$ charges ($0, 2, -2$) are introduced. These contribute to $m_{Z'}^2 = 4\tilde g^2 (v^2_2 + v^2_3)$, with the vev below $4m_W^2/g^2$, implying that the $Z^\prime$ mass cannot be too large.
Therefore, a singlet scalar $S(1,1,0)(1)$ with  vev  $v_s/\sqrt{2}$ is introduced to raise the $Z'$ mass, given by $m^2_{Z^\prime}= \tilde g^2( v_s^2 + 4 v_2^2+4v_3^2)$.
In this case, $Z^\prime$ can have an arbitrarily  mass depending on the values of $v$ and $v_s$. At the electroweak scale $U(1)_{L_\mu-L_\tau}$ $Z^\prime$ has been shown to be allowed by experimental data~\cite{Heeck:2011wj}.

In addition, a discrete exchange symmetry is imposed such that $Z^\prime \to - Z^\prime$,  $H_1 \leftrightarrow H_1$ and $H_2 \leftrightarrow H_3$ with $v_2=v_3=v$.  The unbroken symmetry  forbids the kinetic mixing terms in Eq.~\ref{mix}.
Moreover, it simplifies the Yukawa couplings
 $y_{22}=y_{33}$ and $y_{23}=y_{32}$. Under this symmetry, the $Z^\prime$ interaction and Yukawa terms to leptons are given by
\begin{eqnarray}
L _{H}= &&- \tilde g (\bar l_2 \gamma^\mu L l_2- \bar l_3  \gamma^\mu L l_3 + \bar e_2 \gamma^\mu R e_2 - \bar e_3 \gamma^\mu R e_3) Z^\prime_\mu\nonumber\\
&&- [Y^l_{11} \bar l_1 R e_1 + Y^l_{22} (\bar l_2 R e_2 +\bar l_3 R e_3 ) ] H_1 
-Y^l_{23} (\bar l_2 R e_3 H_2 +\bar l_3 R e_2 H_3 ) + h.c.
\end{eqnarray}
The Yukawa couplings in the second line are non-diagonal, necessitating a transformation to realize the diagonalization.
 The transformation  between  the  lepton mass eigenstates and weak eigenstate basis is given by
\begin{eqnarray}\label{eigen}
\left (
\begin{array}{c}
\mu\\
\tau
\end{array}
\right )
= {1\over \sqrt{2}}
\left (
\begin{array}{rr}
1&\;-1\\
1&\;1
\end{array}
\right )
\left (\begin{array}{c}
e_2\\
e_3
\end{array}
\right ),\quad
\left (
\begin{array}{c}
\nu_{L\mu}\\
\nu_{L\tau}
\end{array}
\right )
= {1\over \sqrt{2}}
\left (
\begin{array}{rr}
1&\;-1\\
1&\;1
\end{array}
\right )
\left (\begin{array}{c}
\nu_{L2}\\
\nu_{L3}
\end{array}
\right ). 
\end{eqnarray}
 In the new basis, the $Z^\prime$ interactions with leptons become the following form as desired,
\begin{eqnarray}
{\cal L}_{Z'}=	- \tilde g (\bar \mu \gamma^\mu \tau +  \bar \tau  \gamma^\mu \mu + \bar \nu_\mu \gamma^\mu L \nu_\tau  + \bar \nu_\tau \gamma^\mu L \nu_\mu) Z'_\mu \;. \label{zprime-changing}
\end{eqnarray}
Therefore, we realize the purpose of converting the flavor-conserving $Z'$ interactions in Eq.~\ref{diagonal} into flavor-changing $Z'$ interactions in Eq.~\ref{zprime-changing}.
Note that the additional neutral and charged Higgs fields from $H_{1,2,3}$ are regarded as   sufficiently heavy such that their corresponding physical effects can be neglected.

\begin{figure*}[!t]
	\centering
	\begin{subfigure}[b]{0.4\textwidth}
		\centering
		\includegraphics[width=\textwidth]{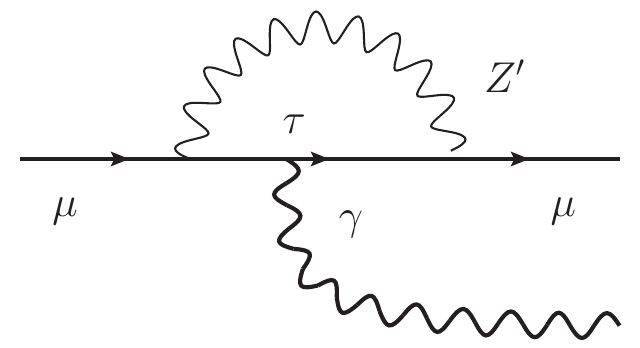}
		\caption{$(g-2)_\mu$.}
		\label{fig:LFVg2}
	\end{subfigure}
	\hfill
	\begin{subfigure}[b]{0.4\textwidth}
		\centering
		\includegraphics[width=\textwidth]{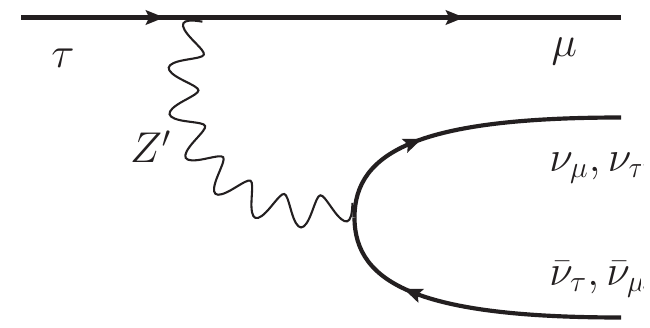}
		\caption{ $\tau \to \mu\nu\nu$.}
		\label{fig:taudecayZ'}
	\end{subfigure}
	\caption{The Feynman diagrams for $(g-2)_\mu$ and $\tau \to \mu\nu\nu$ with maximal flavor-changing $Z'$ interaction.}
	\label{fig:LFV-Zp}
\end{figure*}

The maximal off-diagonal $Z'$ interaction also contributes to $(g-2)_\mu$, with tau leptons running in the one-loop diagram, as shown in Fig.~\ref{fig:LFVg2}.
 The contribution is~\cite{Leveille:1977rc}
\begin{align}
\Delta a_{\mu}=\frac{\tilde g_{Z'}^2 m_{\mu}^2}{4\pi^2}\int^{1}_{0} dx\frac{(x-x^2)(x+\frac{2m_{\tau}}{m_{\mu}}-2)-\frac{(m_{\tau}-m_{\mu})^2}{2m_{Z'}^2}(x^3-x^2(1+\frac{m_{\tau}}{m_{\mu}}))}{m_{\mu}^2 x^2+m_{Z'}^2(1-x)+x(m_{\tau}^2-m_{\mu}^2)}\;.
\end{align}
In the limits of $m_{Z'} \gg m_{\mu}$, it can be simplified as
\begin{align}\label{largeZ'}
\Delta a_{\mu}=&\frac{\tilde g_{Z'}^2 m_{\mu}^2}{4\pi^2}\int_{0}^{1}dx \frac{(x-x^2)(x+\frac{2m_{\tau}}{m_{\mu}})}{m_{Z'}^2(1-x)}
=\frac{\tilde g_{Z'}^2 m_{\mu}^2}{12\pi^2m_{Z'}^2}\left(-2+\frac{3m_{\tau}}{m_{\mu}}\right)\;.
\end{align}
Compared to Eq.~\ref{leading}, there is an enhancement factor $(3 m_\tau/m_\mu-2)$ to $\Delta a_\mu$, which reduces the value ${\tilde g^2/m^2_{Z'}} = (3.96\pm 1.03)\times 10^{-7} \mbox{GeV}^{-2}$. Additionally, the constraint from neutrino trident process donot apply since $Z'$ cannot induce a muon pair $\mu^+\mu^-$.
For small $Z'$ case, we should use the full expression to obtain the corresponding contribution~\cite{Foot:1994vd}.


\begin{figure}[htbp!]
  \centering
  \includegraphics[width=0.8\linewidth]{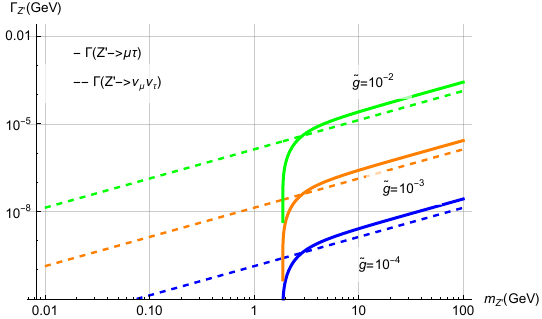}
 \caption{The $Z'$ decay width  with three different parameters, $\tilde g=10^{-4}$(blue), $\tilde g=10^{-3}$(orange) and $\tilde g=10^{-2}$(green). The solid and dashed lines mean $\Gamma(Z'\to\mu^\pm\tau^\mp)$ and $\Gamma(Z'\to \nu_{\mu(\tau)}\bar \nu_{\tau(\mu)})$, respectively.}
 \label{Z'decay}
\end{figure}

The  $Z'$ decay width can be calculated from the maximal flavor-changing interaction,
\begin{eqnarray}
&&\Gamma(Z'\to\mu^\pm\tau^\mp)=\frac{\tilde g_{Z'}^2}{24\pi}\frac{1}{m_{Z'}^5}(m_{Z'}^2-m_{\tau}^2)^2(2m_{Z'}^2+m_{\tau}^2)\;,\notag\\
&& \Gamma(Z'\to \nu_{\mu(\tau)}\bar \nu_{\tau(\mu)})=\frac{g_{Z'}^2m_{Z'}}{24\pi}\;.
\end{eqnarray}
here we ignore the muon mass.
The corresponding decay widths are shown in Fig.~\ref{Z'decay}.
We found that the decay $Z'\to \mu\tau$ only applies for the case of $m_{Z'}\geq m_\mu+m_\tau$.
It becomes apparent that the two decay channels become comparable when $m_{Z'}$ is significantly larger than $m_\mu+m_\tau$.
In the limit of $m_{Z'}>>m_\tau$,  the decay width into neutrinos is half that of the charged lepton decay due to only the left-handed neutrinos,
\begin{eqnarray}
2\mathcal{B}(Z^{\prime}\to\nu_{\mu(\tau)}\bar{\nu}_{\tau(\mu)})=\mathcal{B}(Z^{\prime}\to \mu^{\pm}\tau^{\mp}).
\end{eqnarray}
This corresponds to $Br(Z'\to\nu_{\mu(\tau)}\bar{\nu}_{\tau(\mu)})=1/3$ or 1, which depends on whether the process $Z'\to \mu\tau$ can happen or not.

The above maximal off-diagonal $Z'$ interaction in Eq.~\ref{zprime-changing} can induce the decay $\tau\to \mu\nu\bar\nu$ with off-shell or on-shell $Z'$ boson as shown in Fig.~\ref{fig:taudecayZ'}.
For the case of  $m_{Z'} \geq m_{\tau}+m_{\mu}$, it  corresponds to  $\mathcal{B}r(Z'\to \nu_{\mu}\nu_{\tau})=1/3$. In this case, the main experimental constraint is from the decay 
$\tau \to \mu \bar \nu \nu$ by the off-shell $Z'$ boson.  
Including the SM $W$ and the new $Z^\prime$ contributions,  the effective Lagrangian $ L_{W+ Z^\prime}$ as
\begin{eqnarray}\label{SM}
\mathcal{L}_{W+ Z^\prime}=-{g^2\over 2 m^2_{W}}  \bar \nu_\tau \gamma^\mu L \nu_\mu \bar \mu \gamma_\mu L \tau  + {\tilde g^2\over {m_\tau^2- m^2_{Z^\prime}}}(\bar \nu_\tau \gamma^\mu L \nu_\mu + \bar \nu_\mu \gamma^\mu L \nu_\tau)\bar \mu \gamma_\mu  \tau\;,
\end{eqnarray}
here we choose $q^2=m_\tau^2$. And we consider  the family
lepton-number preserving three-body decay $\tau\to \mu \bar \nu_\mu \nu_\tau$ and the family lepton number violating decay $\tau\to \mu \bar \nu_\tau \nu_\mu$ at the same time since the neutrino flavor of final states is indistinguishable in the experimental measurement.

The current experimental measurements are in terms of the ratio as 
\begin{eqnarray}
R^\tau_{\mu e}= \frac{\Gamma(\tau^-\to \mu^-\nu_\tau \bar \nu_\mu)}{\Gamma(\tau^-\to e^-\nu_\tau \bar \nu_e)}\;.
\end{eqnarray}
The corresponding  SM predictions~\cite{Pich:2009zza} and experimental vaules~\cite{HFLAV:2022esi} from HFLAV collaboration global fit
\begin{eqnarray}
R^\tau_{\mu e}|_{SM}= 0.972564\pm0.00001,\;\;\; R^\tau_{\mu e}|_{exp}= 0.9761\pm 0.0028,\;.
\end{eqnarray}
Recently, Belle-II updates the result as $R^\tau_{\mu e}|_{Belle}=0.9675\pm0.0007\pm0.0036$~\cite{Corona:2024nnn}. 
The above two experimental results can be combined into the average weighted  value $R^\tau_{\mu e}|_{aver}=0.972968\pm 0.002233$.
Because our model does not affect $\tau \to e \nu_\tau \bar \nu_e$, the ratio can be transformed as 
\begin{eqnarray}
 R^\tau_{exp}&\equiv&\frac{\Gamma(\tau^-\to\mu^-\overline{\nu}_\mu\nu_\tau)}{\Gamma(\tau^-\to\mu^-\overline{\nu}_\mu\nu_\tau)_{\mathrm{SM}}}
 =1.00042\pm 0.00230\;.
\end{eqnarray}
The interaction in Eq.~\ref{SM} will contribute the decay $\tau\to \mu \nu\bar\nu$ as~\cite{Foot:1994vd,Altmannshofer:2016brv} 
\begin{eqnarray}\label{Z'tau}
R^\tau= 1+\frac{\tilde g^2}{g^2}\frac{4m_W^2}{m_{Z^{\prime}}^2}f\left(\frac{m_\tau^2}{m_{Z^{\prime}}^2}\right)+\frac{\tilde g^4}{g^4}\frac{4m_W^4}{m_{Z^{\prime}}^4}g\left(\frac{m_\tau^2}{m_{Z^{\prime}}^2}\right)
\end{eqnarray}
with the functions
\begin{align}
&f(z) =\frac2{z^4}\left[\frac56z^3+2z^2-2z-(1-z)^2(2+z)\log(1-z)\right]\mathrm{~,} \\
&g(z) =\frac2{z^4}\left[-z^3-3z^2+6z+6(1-z)\log(1-z)\right]\mathrm{~.} 
\end{align}
here we consider the momentum transfer for $m_{Z'}$ of the order of the tau mass.
In the limit of $m_{Z'}>>m_\tau$, we can obtain $\lim_{z\to0}f(z)=1$ and $\lim_{z\to0}g(z)=1$.

Further, $R^\tau$ can constrain the model parameters.
Unfortunately, 
data from $R^{\tau}=1.00042\pm0.00230$, 
excludes the total parameter regions by more than   5 $\sigma$ level  if required to solving the $(g-2)_\mu$ anomaly simultaneously.

For the case of $ m_{Z'} \leq m_{\tau}-m_{\mu}$, the main process is the tau invisible decay of $\tau \to \mu Z'$ due to the kinetic allowed requirement. The corresponding decay width is
\begin{align}
\Gamma(\tau \to\mu Z')=\frac{\tilde g_{Z'}^2 m_{\tau}^3}{16\pi m_{Z'}^2}\left(1+\frac{2m_{Z'}^2}{m_{\tau}^2}\right)\left(1-\frac{m_{Z'}^2}{m_{\tau}^2}\right)^2\;.
\end{align}
here the muon mass is neglected.
The current experimental upper limit is $Br(\tau\to \mu Z')<6\times 10^{-4}$ at 95\%CL~\cite{ParticleDataGroup:2022pth}, which can constrain our model parameters.
Besides, the on-shell $Z'$ can further decay into $ \nu_{\mu(\tau)}\bar \nu_{\tau(\mu)}$ with 
 the branching ratio equal to one.
In this case, the constraint from $Br(\tau \to \mu Z')$ is much stronger than  $R^\tau$.

\begin{figure}[htbp!]
  \centering
  \includegraphics[width=0.6\linewidth]{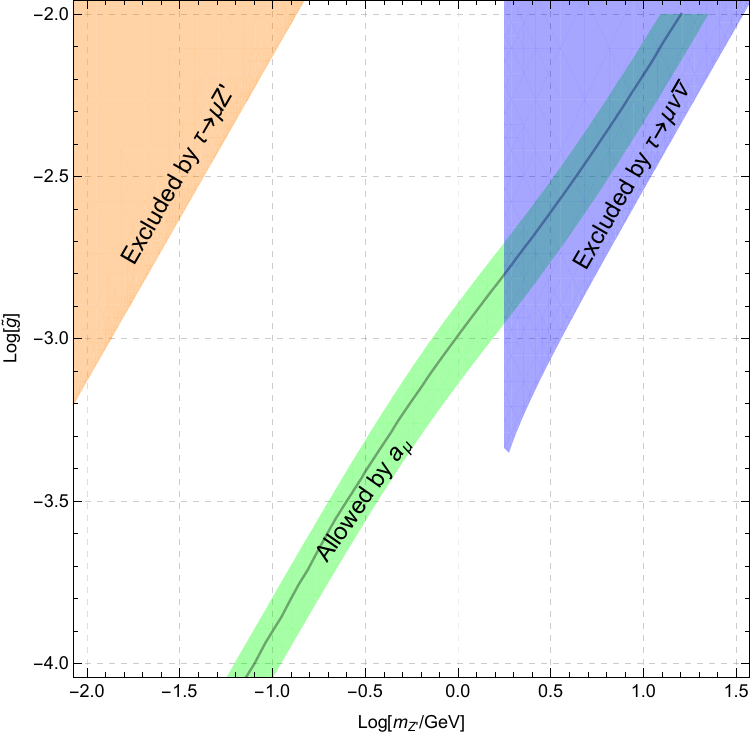}
 \caption{The allowed parameter regions for the maximal flavor-changing $Z'$ model. The green region means the $(g-2)_\mu$ allowed ranges within 2$\sigma$ errors and black line represents the central value of $a_\mu=1.81\times 10^{-9}$. 
 For small $m_{Z'}$ region, the decay $\tau\to\mu Z'$  excludes the region in orange.
 For  $m_{Z'}>m_\tau-m_\mu$, the decay $\tau\to\mu \nu\bar\nu$  excludes the region in blue.}
 \label{Z'region}
\end{figure}

By combining the constraints from $(g-2)_\mu$, $Br(\tau \to \mu Z')$, and $R^\tau$, we plot the corresponding parameter regions, as illustrated in Fig.~\ref{Z'region}.
The black line and green region represent the allowed parameter ranges for $(g-2)_\mu$, corresponding to the central value and within the 2$\sigma$ range.
The orange and blue regions mean the excluded areas due to the decays $\tau \to \mu Z'$ and $\tau \to \mu \nu \bar{\nu}$, respectively.
We find that the decay $\tau \to \mu Z'$ only applies in the case where $m_{Z'} < m_\tau - m_\mu$.
The most stringent constraints arise from the decay $\tau \to \mu \nu \bar{\nu}$, which exclude the entire parameter space for $m_{Z'} > m_\tau - m_\mu$.
This indicates that the $a_\mu$ allowed region only exists in the case of small $m_{Z'}$.
In order to avoid the stringent bounds, we attempt to introduce new interactions  involving singly-charged scalar in the following section.

\section{New effects induced by Singly-Charged Higgs}

In the previous section, we identified that the most stringent constraint on the interaction arises from the decay $\tau \to \mu \nu \bar{\nu}$, which is induced by the maximal off-diagonal $Z'$ neutral current.
In order to weaken this constraints, we attempt to introduce new interactions. 
Analogous to the neutral and charged currents in SM, we propose introducing a singly-charged scalar that can mediate the process $\tau \to \mu \nu \bar{\nu}$ through charged current interactions.

In general, introducing  singly-charged scalars can lead to three different scenarios: a weak singlet scalar (Case I), a triplet scalar (Case II), and a doublet scalar (Case III).
The corresponding operator is described as
\begin{align}\label{operator}
\mathcal{L}_{h^\pm}=C_{LL}\bar{\nu}_{La}e^{c}_{Lb}h^{-}+C_{LR}\bar{\nu}_{La}e_{Rb}h^{+} +h.c.\;.
\end{align}
In the context of renormalizable theories, the LL operator in the first term arises from gauge-invariant leptonic interactions involving either a weak singlet or a triplet, while the LR operator in the second term is realized when the interaction involves a doublet.
Furthermore, both operators can also be generated within effective field theories where the heavy states have been integrated out.

For the singly-charged Higgs, the contribution to lepton $(g-2)_l$ is expressed as~\cite{Leveille:1977rc}
\begin{eqnarray}
a_{l}=&\frac{-q_{h^+}m_{l}^2}{8\pi^2}\int_{0}^{1}dx \frac{C_{s}^2\left(x^3-x^2+\frac{m_{F}}{m_{l}}(x^2-x)\right)+C_{p}^2\left(x^3-x^2-\frac{m_{F}}{m_{l}}(x^2-x)\right)}{m_{l}^2x^2+(m_{h^+}^2-m_{l}^2)x+m_{F}^2(1-x)}\;.
\end{eqnarray}
here $m_F$ means the fermion mass running in the loop. $q_{h^+}$ means the carried electric charge in unit of positron e.
The coupling is defined as  $C_{s}=\frac{C_{LL}+C_{LR}}{2}$ and $C_{p}=\frac{C_{LR}-C_{LL}}{2}$. 
If only one term in Eq.~\ref{operator} is present, regardless of whether it is Case I, II, or III, it will lead to
 $C_s^2=C_p^2=C_{LL(LR)}^2/4$.

For the large $m_{h^+}$ with hundreds of GeV as required by the current collider constraints, the expression can be further simplified to
\begin{eqnarray}\label{chargedg2}
a_{l}=&\frac{m_{l}^2}{32\pi^2}C_{LL(LR)}^2\int_{0}^{1}dx \frac{2\left(x^3-x^2\right)}{m_{h^+}^2 x}
=-\frac{1}{96\pi^2}\frac{m_{l}^2}{m_{h^{+}}^2}C_{LL(LR)}^2\;.
\end{eqnarray}
We found that the singly-charged Higgs results in a negative contribution to $(g-2)_l$, which is opposite to the positive contribution from maximal off-diagonal $Z'$ interactions.

Note that in Eq.~\ref{chargedg2}, we retain the general expression for charged fermions, as the singly-charged Higgs can contribute to $(g-2)_l$ for fermions of other generations.
For the experimental measurement of $(g-2)_\mu$, it has been discussed in Eq.~\ref{muong2}.
Regarding the case of $(g-2)_e$, the current status remains unclear, as the most precise measurements of the fine structure constant $\alpha$ are not in agreement.
The latest measurement of $(g-2)_e$~\cite{Fan:2022eto} is in tension with the SM prediction by either $+2.2\sigma$ or $-3.7\sigma$\cite{Davoudiasl:2023huk}, depending on whether the fine-structure constant $\alpha$ obtained from Rb~\cite{Morel:2020dww} or Cs~\cite{Parker:2018vye} is used as input
\begin{eqnarray}
&&rubidium: \;\;\;\Delta a_e(Rb)=(34+16)\times 10^{-14},\;\;2.2\sigma\;,\nonumber\\
&&cesium:\;\;\;\Delta a_e(Cs)=(-101+27)\times 10^{-14},\;\;-3.7\sigma\;,
\end{eqnarray}
Due to the negative contribution from the singly-charged Higgs, we use the experimental values from cesium, $\Delta a_e(\text{Cs})$, for the analysis that follows.
For $(g-2)_\tau$, the current experimental values have large uncertainties: 
$a_\tau=0.0009_{-0.0031}^{+0.0032}$(CMS~\cite{CMS:2024skm}), with the SM prediction $a_\tau^{SM}=(1.17721\pm0.00005)\times10^{-3}$~\cite{Eidelman:2007sb}. 
Therefore, we do not consider $(g-2)_\tau$ in our analysis.

In this section, we will analyze the possibility that whether the singly-charged Higgs can reconcile the discrepancy from maximal flavor-violating $Z'$ interactions between $(g-2)_l$ and $\tau \to \mu \nu \bar{\nu}$ in the three different cases mentioned above.
In order to maintain the discrete exchange symmetry $1 \leftrightarrow 1$ and $2 \leftrightarrow 3$, we introduce three new scalars with $U(1)_{L_\mu - L_\tau}$ charges of (0, 2, -2), respectively.

\subsection{ singly-charged Higgs from weak singlet (Case I) }

For the LL operator in Eq.~\ref{operator}, there are two distinct contributions, generated by singlet and triplet scalars.
Firstly, we focus on the case of singly-charged Higgs from  weak singlet.
In this case, the Yukawa interaction with lepton pairs can be written directly as 
\begin{align}\label{singletL}
\mathcal{L}_{h^{+}}&=y_{ab}\overline{L^{c}_{La}}i \sigma^2 L_{Lb}h^{+}+h.c.
=y_{ab}\begin{pmatrix} \overline{\nu^{c}_{La}},& \overline{e^{c}_{La}} \end{pmatrix}\begin{pmatrix} 0 &1\\-1&0\end{pmatrix} \begin{pmatrix} \nu_{Lb} \\ e_{Lb}\end{pmatrix} h^{+}+h.c.\\ \nonumber
&=y_{ab}(\overline{\nu^{c}_{La}} e_{Lb}h^{+}-\overline{e^{c}_{La}} \nu_{Lb}h^{+})+h.c.
\end{align}
here  $L_L$ denotes the lepton doublet  and $\sigma^2$ means the pauli matrix. The existence of Clebsh factor $i\sigma^2$ generates the anti-symmetric feature for the Yukawa coupling $y_{ab}$. The anti-symmetry makes $y_{aa}=0$, which results in the disappearance of the same lepton flavors.

Due to the definition $\psi^c=c \bar\psi^T$ with charge conjugation $c$ and fermion $\psi$, one can naturally prove the relations of $\overline{\psi_{i}^{c}}P_{L}\psi_{j}=\overline{\psi_{j}^{c}}P_{L}\psi_{i}$. Therefore,  Eq.~\ref{singletL} can be expressed as 
\begin{align}
\mathcal{L}_{h^{+}}=y_{ab}( \overline{\nu^{c}_{La}} e_{Lb}-\overline{\nu^{c}_{Lb}} e_{La})h^{+}+h.c.\;.
\end{align}

In order to maintain the discrete symmetry 
$1 \leftrightarrow 1$ and $2 \leftrightarrow 3$, we introduce three different singly-charged singlet Higgs $h^+_{1,2,3}$ with the corresponding $U(1)_{L_\mu-L_\tau}$ charges (0,2,-2), respectively.
The antisymmetric Yukawa coupling cancels the contributions from $h^\pm_{2,3}$, leaving only the effect from $h^\pm_{1}$, with the interaction given by
\begin{align}\label{singleth23}
\mathcal{L}_{h^{+}}&=2y_{23}( \overline{\nu^{c}_{L2}} e_{L3}-\overline{\nu^{c}_{L2}} e_{L3})h_1^{+}+h.c.\nonumber\\
&=y_{23}\begin{pmatrix}  \overline{\nu^{c}_{L\mu}} &  \overline{\mu^{c}_{L\tau}}\end{pmatrix} \begin{pmatrix} 1&-1\\1&1\end{pmatrix} \begin{pmatrix} 0 &y_{23}h_1^{+}\\ -y_{23}h_1^{+}&0\end{pmatrix} \begin{pmatrix} 1&1\\-1&1\end{pmatrix} \begin{pmatrix} \nu_{L\tau}\\  \tau_{L\tau}\end{pmatrix} +h.c.\\ \nonumber
&=y_{23}h_{1}^{+}(- \overline{\mu^{c}_{L}}\nu_{\tau L}+ \overline{\nu^{c}_{L\mu}}\tau_{L}-\overline{\nu^{c}_{L\tau}}\mu_{L}+\overline{\tau^{c}_{L}}\nu_{L\mu})+h.c.\;,
\end{align}
here in the second line we use the transformation in Eq.~\ref{eigen} between the weak eigen-states and mass  eigen-states.

From Eq.~\ref{singleth23}, we find that $h_1^\pm$ mediates only flavor-changing interactions. Consequently, it does not produce $\mu^+\mu^-$ pairs in the neutrino trident process, and thus $h_1^\pm$ does not contribute to this process.

\begin{figure}[htbp!]
  \centering
  \includegraphics[width=0.8\linewidth]{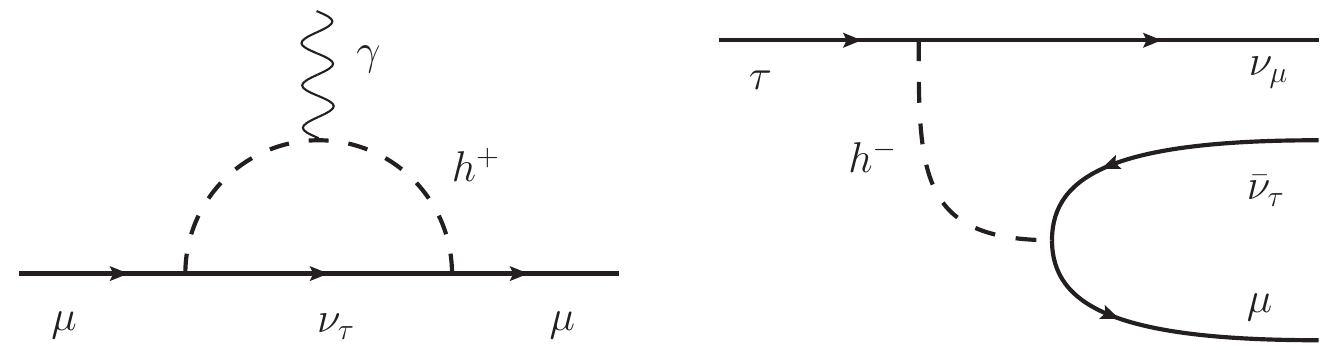}
 \caption{The corresponding Feymann diagram contribution for singly-charged Higgs from weak singlet. 
The left panel means the muon $(g-2)_\mu$ mediated by $h^+_1$.  
The right panel means the decay process $\tau\to \mu \nu_\mu \bar \nu_\tau$ mediated by $h^-_1$.}
 \label{singletg-2}
\end{figure}

The interaction described in Eq.~\ref{singleth23} contributes to $(g-2)_\mu$ at the one-loop level, as shown in the left panel of Fig.~\ref{singletg-2}.
By using the general expression in Eq.~\ref{chargedg2} with $l = \mu$, one can obtain the contribution to $(g-2)_\mu$ in the limit of large $m_{h^+_1}$,
\begin{align}
a_{\mu}=-\frac{1}{96\pi^2}\frac{m_{\mu}^2}{m_{h_1^{+}}^2}4y_{23}^2=-\frac{1}{24\pi^2}\frac{m_{\mu}^2}{m_{h_1^{+}}^2}y_{23}^2\;,
\end{align}
here we use the relation $\overline{\mu^{c}_{L}}\nu_{\tau L}=\overline{\tau^{c}_{L}}\nu_{\mu L}$ to obtain the 4 times enhancement.

Similarly, the $h_1^-$ couplings in Eq.~\ref{singleth23} can induce the decay  $\tau\to \mu \nu\bar{\nu}$ at the tree level with the amplitude as shown in the right panel of Fig.~\ref{singletg-2},
\begin{eqnarray}
   \mathcal{L}_{h_1^-}= -\frac{4y_{23}^2}{m_{h^+_1}^2}\bar \nu^c_{\mu_L} \tau_L \bar \mu_L \nu_\tau^c
    =-\frac{2y_{23}^2}{m_{h^+_1}^2}\bar \mu_L \gamma^\mu P_L \tau_L \bar \nu_\tau \gamma_\mu P_L \nu_\mu\;.
\end{eqnarray}
here in the second equation we use the fierz transformation and $\overline{\nu_i^c}\gamma^\mu P_R \nu_j^c=-\bar \nu_j \gamma^\mu P_L \nu_i$.

Combining with the contribution in Eq.~\ref{SM} from  W and $Z'$ boson, we can obtain the total amplitude $\mathcal{M}_{total}$ as  
 \begin{align}\label{taudecaysinglet}
 \mathcal{M}_{total}(\tau\to \mu \nu\bar{\nu})=&\left(-\frac{g^2}{4m_{W}^2}+\frac{\tilde g^2}{m_\tau^2-m_{Z'}^2}-\frac{y_{23}^2}{m_{h_{1}^{+}}^2}\right)\bar{\mu}\gamma^{\mu}\tau\bar{\nu}_{\tau}\gamma_{\mu}P_{L}\nu_{\mu}\\ \nonumber
 &-\left(-\frac{g^2}{4m_{W}^2}-\frac{y_{23}^2}{m_{h_{1}^{+}}^2}\right)\bar{\mu}\gamma^{\mu}\gamma_{5}\tau\bar{\nu}_{\tau}\gamma_{\mu}P_{L}\nu_{\mu}-\frac{\tilde g^2}{m_{Z'}^2}\bar{\mu}\gamma^{\mu}\tau\bar{\nu}_{\mu}\gamma_{\mu}P_{L}\nu_{\tau}\;,
 \end{align}
Note that the total amplitude indicates that the new charged singlet will further enhance the effects of the SM W boson and $Z'$ boson for large $m_{Z'}$ due to their same alignment in sign. This exacerbates the discrepancy between $(g-2)_\mu$ and $\tau \to \mu \nu \bar{\nu}$.
Consequently, this will affect the ratio $R^\tau$ as
\begin{eqnarray}
  R^\tau=  \left(1+\frac{4m_W^2}{g^2}\frac{y_{23}^2}{m_{h_1^+}^2}\right)^2+\frac{\tilde g^2}{g^2}\frac{4m_W^2}{m_{Z^{\prime}}^2}\left(1+\frac{4m_W^2}{g^2}\frac{y_{23}^2}{m_{h_1^+}^2}
  \right)f\left(\frac{m_\tau^2}{m_{Z^{\prime}}^2}\right)+\frac{\tilde g^4}{g^4}\frac{4m_W^4}{m_{Z^{\prime}}^4}g\left(\frac{m_\tau^2}{m_{Z^{\prime}}^2}\right)\;.
\end{eqnarray}
The presence of the term $y_{23}^2/m_{h_1^+}^2$ further increases the predicted value of $R^\tau$, which is excluded by the experimental value $R^\tau_{exp}$.

\begin{figure*}[!t]
	\centering
	\begin{subfigure}[b]{0.48\textwidth}
		\centering
		\includegraphics[width=\textwidth]{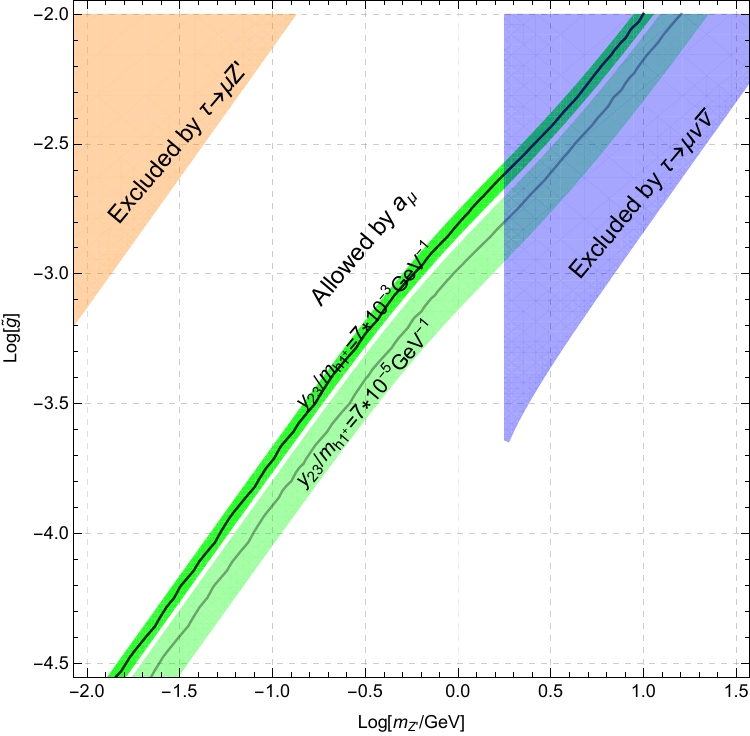}
		\caption{}
		\label{fig:singlet2}
	\end{subfigure}
	\hfill
	\begin{subfigure}[b]{0.48\textwidth}
		\centering
		\includegraphics[width=\textwidth]{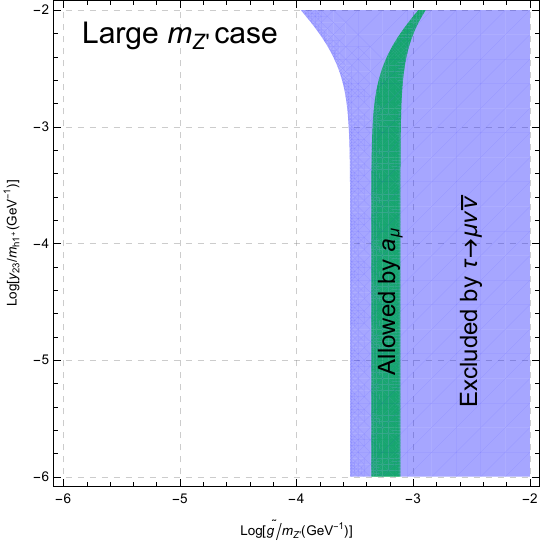}
		\caption{}
		\label{fig:singlet1}
	\end{subfigure}
	\caption{The allowed parameter regions for the maximal flavor-changing $Z'$ models with singly-charged Higgs from weak singlet (Case I). The black line and green region represent the $(g-2)_\mu$ allowed ranges for the central value of $a_\mu=1.81\times 10^{-9}$ and within 2$\sigma$ errors, respectively. For small $m_{Z'}$ region, the decay $\tau\to\mu Z'$ excludes the region in orange. For $m_{Z'}>m_\tau-m_\mu$, the decay $\tau\to\mu \nu\bar\nu$ excludes the region in blue. a) shows the $\log[m_{Z'}/\text{GeV}]-\log[\tilde g]$ plane with two different charged scalar parameters, $y_{23}/m_{h_1^+}=7\times 10^{-3}$GeV$^{-1}$ and $y_{23}/m_{h_1^+}=7\times 10^{-5}$GeV$^{-1}$. b) shows the $\log[\tilde g/m_{Z'}(\text{GeV}^{-1})]-\log[y_{23}/m_{h_1^+}(\text{GeV}^{-1})]$ plane for the large $m_{Z'}$ case.}
	\label{fig:singlet}
\end{figure*}

Based on the above analysis, we plot the corresponding parameter regions as shown in Fig.~\ref{fig:singlet}.
The black line and green region represent the $(g-2)_\mu$ allowed ranges: the black line indicates the central value of $a_\mu = 1.81 \times 10^{-9}$, while the green region represents the range within 2$\sigma$ errors.
For the region of small $m_{Z'}$, the decay $\tau \to \mu Z'$ excludes the orange region.
For $m_{Z'} > m_\tau - m_\mu$, the decay $\tau \to \mu \nu \bar{\nu}$ excludes the blue region.
In Fig.~\ref{fig:singlet2}, we illustrate the effect of two different charged Higgs parameters, $y_{23}/m_{h_1^+}=7\times 10^{-3}$GeV$^{-1}$ and $y_{23}/m_{h_1^+}=7\times 10^{-5}$GeV$^{-1}$, on the results.
We find that as $y_{23}/m_{h_1^+}$ increases, the allowed region for $a_\mu$ shifts upwards and becomes narrower. This occurs because the opposing contributions from the maximal flavor-changing $Z'$ and the singly-charged Higgs $h_1^+$ compete with each other.
In Fig.~\ref{fig:singlet1}, we illustrate the case of large $m_{Z'}$ using the formula in Eq.~\ref{largeZ'}. This indicates that the allowed region for $(g-2)_\mu$ is entirely excluded by the decay $\tau \to \mu \nu \bar{\nu}$.

 Therefore, we find that the singly-charged Higgs $h_1^+$ from a weak singlet cannot reconcile the discrepancy between $(g-2)_\mu$ and $\tau \to \mu \nu \bar{\nu}$ in the large $m_{Z'}$ case.
 Furthermore, the introduction of $h_1^+$ will increase the $Z'$ coupling parameter $\tilde{g}$ by a factor of approximately 2 if it is required to explain the positive experimental value $a_\mu^{exp}$.

\subsection{ singly-charged Higgs from weak triplet (Case II) }

In this part, we focus on another  LL operator in Eq.~\ref{operator}   that is induced by a weak triplet scalar.
The corresponding Yukawa coupling term can be expressed by 
\begin{align}
\mathcal{L}&=y_{ab}\begin{pmatrix}  \overline{\nu^{c}_{La}}, &  \overline{e^{c}_{La}}\end{pmatrix} i\sigma^{2} \begin{pmatrix}\frac{\Delta^{+}}{\sqrt{2}}&\Delta^{++} \\ \Delta^{0} & -\frac{\Delta^{+}}{\sqrt{2}}\end{pmatrix}  \begin{pmatrix} \nu_{Lb}\\  e_{Lb}\end{pmatrix}  +h.c. \\ \nonumber 
&=y_{ab}\left(\overline{\nu_{La}^c}\Delta^{0}\nu_{Lb}-\frac{\Delta^{+}}{\sqrt{2}}(\overline{e_{La}^{c}}\nu_{Lb}+\overline{\nu_{La}^{c}}e_{Lb})-\overline{e_{La}^{c}}e_{Lb}\Delta^{++}\right)+h.c.\;,
\end{align}
By focusing solely on the singly-charged Higgs interaction, we find a positive sign of $\overline{e_{La}^{c}} \nu_{Lb} + \overline{\nu_{La}^{c}} e_{Lb}$, which differs significantly from the negative contribution in the singlet case.
This indicates that for the same flavor ($a=b$), the Yukawa couplings do not vanish, 
$\mathcal{L}=-\sqrt{2}y_{aa}\overline{\nu_{La}^{c}}e_{La}\Delta^{+}+h.c.$.

Similarly, three triplets $\Delta_{1}(0)$, $\Delta_{2}(2)$, and $\Delta_{3}(-2)$ are introduced to ensure the discrete exchange symmetry $1 \leftrightarrow 1$ and $2 \leftrightarrow 3$. Here the numbers in parentheses denote the corresponding $U(1)_{L_{\mu}-L_{\tau}}$ charges.
This discrete symmetry also implies $m_{\Delta_2^+}=m_{\Delta_3^+}$.
Therefore, the Yukawa coupling terms can be expressed as
\begin{align}\label{triplet}
\mathcal{L}=-\frac{\sqrt{2}}{2}\begin{pmatrix} \overline{\nu_{e}^{c}}, \overline{\nu_{\mu}^{c}}, \overline{\nu_{\tau}^{c}}\end{pmatrix} \begin{pmatrix} 2y_{11}\Delta^+_{1} &0&0\\0&y_{22}\Delta_{2+3}^{+}-2y_{23}\Delta^+_{1}&y_{22}\Delta_{2-3}^{+}\\0&y_{22}\Delta_{2-3}^{+}&y_{22}\Delta_{2+3}^{+}+2y_{23}\Delta_{1}^+\end{pmatrix}\begin{pmatrix} e_L\\ \mu_L \\ \tau_L\end{pmatrix}+h.c,\;\nonumber\\
-\frac{1}{2}\begin{pmatrix} \overline{e^{c}}, \overline{\mu^{c}}, \overline{\tau}^{c}\end{pmatrix} \begin{pmatrix} 2y_{11}\Delta^{++}_{1} &0&0\\0&y_{22}\Delta_{2+3}^{++}-2y_{23}\Delta^{++}_{1}&y_{22}\Delta_{2-3}^{++}\\0&y_{22}\Delta_{2-3}^{++}&y_{22}\Delta_{2+3}^{++}+2y_{23}\Delta_{1}^{++}\end{pmatrix}\begin{pmatrix} e_L\\ \mu_L \\ \tau_L\end{pmatrix}+h.c
\end{align}
where $\Delta_{2+3}^{+,++}$ and $\Delta_{2-3}^{+,++}$ represent $\Delta_{2}^{+,++}+\Delta_{3}^{+,++}$ and $\Delta_{2}^{+,++}-\Delta_{3}^{+,++}$, respectively.
Here, we use the transformation given in Eq.~\ref{eigen} to relate the weak eigenstates to the mass eigenstates.
Note that $\Delta_{1}^+$ mediates only flavor-conserving interactions, whereas $\Delta_{2,3}^+$ can mediate both flavor-conserving and flavor-changing interactions between tau and muon leptons.
The different signs in the diagonal and off-diagonal terms for $\Delta^+_{2,3}$ cancel each other out, as seen in processes such as $\tau \to \mu \gamma$.

Note that we also include the couplings for doubly-charged Higgs $\Delta^{++}_i$ since they could affect the certain physical processes. As mentioned in Appendix, the mass differences between singly-charged and doubly-charged scalars  depend on the vacuum expectation value $\tilde v_i$, which is  constrained by EW $\rho $ parameter to satisfy $\tilde v_i<O(1)$GeV~\cite{ParticleDataGroup:2022pth}. Consequently, the masses for $\Delta^+_i$ and $\Delta^{++}_i$ are approximately degenerate, necessitating the consideration of their corresponding physical effects.
Current collider constraints place the lower bound on the doubly-charged scalar mass at approximately  800GeV~\cite{ATLAS:2017xqs} (84GeV~\cite{Kanemura:2014ipa}), depending on the dominant decay
modes $\Delta^{++}\to l^+ l^+$ ($\Delta^{++}\to W^+ W^+$).  For the degenerate case with $\tilde v_i\sim$O(GeV), collider constraints have already excluded doubly-charged scalar masses below 420GeV~\cite{Ashanujjaman:2021txz}.

\begin{figure}[htbp!]
  \begin{minipage}[t]{0.8\linewidth}
  \centering
  \includegraphics[width=1\linewidth]{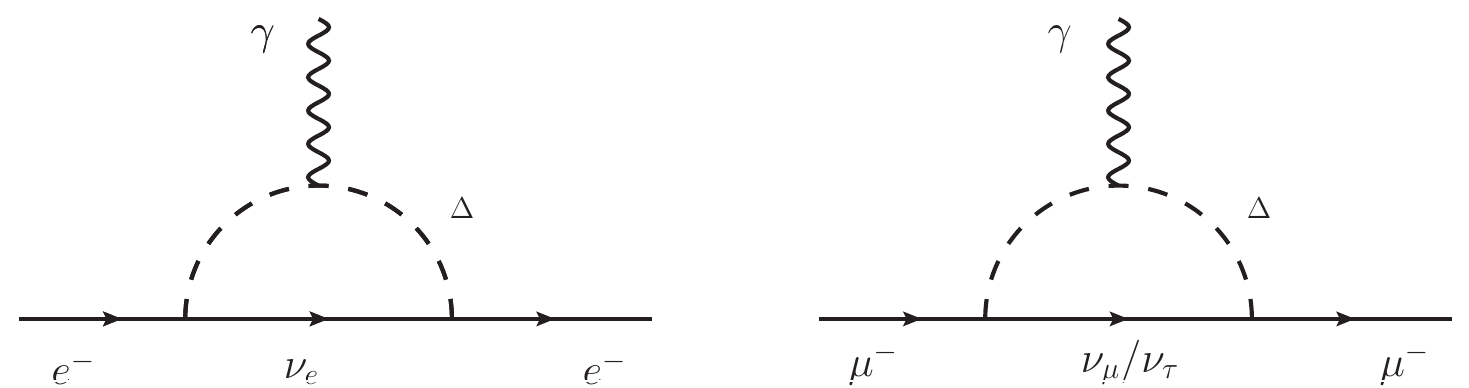}
  \includegraphics[width=0.5\linewidth]{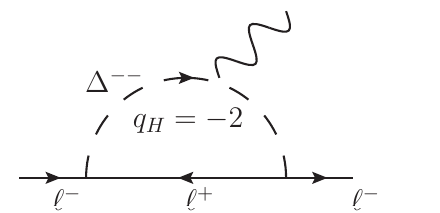}
  \includegraphics[width=0.49\linewidth]{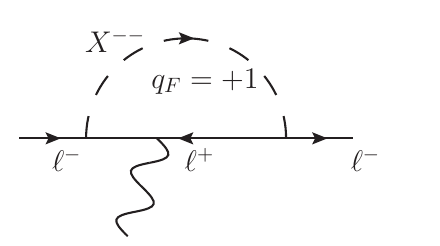}
    \end{minipage}
\caption{The electron $(g-2)_e$ and muon $(g-2)_\mu$  for singly-charged and doubly-charged Higgs from weak triplet. }
 \label{tripletg-2}
\end{figure}

Additionally, the doubly-charged scalars will contribute to lepton $(g-2)_l$ as 
shown in Fig.~\ref{tripletg-2} with total effects as~\cite{Leveille:1977rc}
\begin{eqnarray}\label{doublyg-2}
    a_l=-\frac{2m_l^2}{96\pi^2}\frac{C_{LL}^2}{m^2_{\Delta^{++}}} -\frac{m_l^2}{48\pi^2}\frac{C_{LL}^2}{m^2_{\Delta^{++}}}
    =-\frac{m_l^2}{24\pi^2}\frac{C_{LL}^2}{m^2_{\Delta^{++}}}\;,
\end{eqnarray}
here the first and second term corresponds to the lower left and lower right panel in Fig.~\ref{tripletg-2}, respectively.

Besides, the $\Delta^+_1$ can generate the  electron $(g-2)_e$ as shown in the upper left panel of Fig.~\ref{tripletg-2}.
By choosing $l=e$ in the general expression in Eq.~\ref{chargedg2} and combining the effects   from doubly-charged scalars in Eq.~\ref{doublyg-2}, we obtain the total contribution as 
\begin{eqnarray}
 a_{e}=-\frac{m_{e}^2}{96\pi^2}\frac{2y_{11}^2}{m_{\Delta_1^{+}}^2} -\frac{m_{e}^2}{24\pi^2}\frac{y_{11}^2}{m_{\Delta_1^{++}}^2}
 =-\frac{y_{11}^2}{16\pi^2}\frac{m_{e}^2}{m_{\Delta_1^{+}}^2}\;,
\end{eqnarray}
here $2y_{11}^2$ in the first equation comes from the coupling coefficient $-\sqrt{2}y_{1}$. And the second equation uses the degenerate condition $m_{\Delta_i^+}=m_{\Delta_i^{++}}$
The experimental data $a_e^{exp}(Cs)$ constrains $y^2_{11}/m^2_{\Delta^+_1}=(0.61\pm0.16)\times 10^{-3}$ GeV$^{-2}$.

For muon $(g-2)_\mu$, it can  be induced by $\Delta_{1,2,3}$ simultaneously as shown in the upper right panel of Fig.~\ref{tripletg-2}. The corresponding total contributions with singly- and doubly-charged scalars are
\begin{eqnarray}
   a_{\mu}&=&-\frac{1}{96\pi^2}\frac{m_{\mu}^2}{m_{\Delta_1^{+}}^2}2y_{23}^2  
   -\frac{1}{96\pi^2}m_{\mu}^2
   \left(\frac{y_{22}^2}{2m_{\Delta_2^{+}}^2}
   +\frac{y_{22}^2}{2m_{\Delta_3^{+}}^2}\right)\times 2
   -\frac{m_{\mu}^2}{24\pi^2}\frac{y_{23}^2}{m_{\Delta_1^{++}}^2}  
   -\frac{m_{\mu}^2}{24\pi^2}
   \left(\frac{y_{22}^2}{2m_{\Delta_2^{++}}^2}
   +\frac{y_{22}^2}{2m_{\Delta_3^{++}}^2}\right)\times 2
   \nonumber\\
   &=& -\frac{m_{\mu}^2}{16\pi^2}\frac{y_{23}^2}{m_{\Delta_1^{+}}^2}  
   -\frac{m_{\mu}^2}{16\pi^2}
   \frac{y_{22}^2}{m_{\Delta_2^{+}}^2}\;,
\end{eqnarray}
here the factor $2y_{23}^2$ in the first term and $y_{22}^2/2$ in the second term come from the coupling coefficient $\sqrt{2}y_{23}$ and $-y_{22}/\sqrt{2}$. And the additional factor 2 in the second term comes from the neutrino flavors $(\nu_\mu,\nu_\tau)$ running in the loop.
Similar situation also applies to the case for doubly-charged scalars.
The second equation uses the relations $m_{\Delta^{+,++}_2}=m_{\Delta^{+,++}_3}$  to simplify the results.

\begin{figure}[htbp!]
  \begin{minipage}[t]{0.8\linewidth}
  \centering
  \includegraphics[width=1\linewidth]{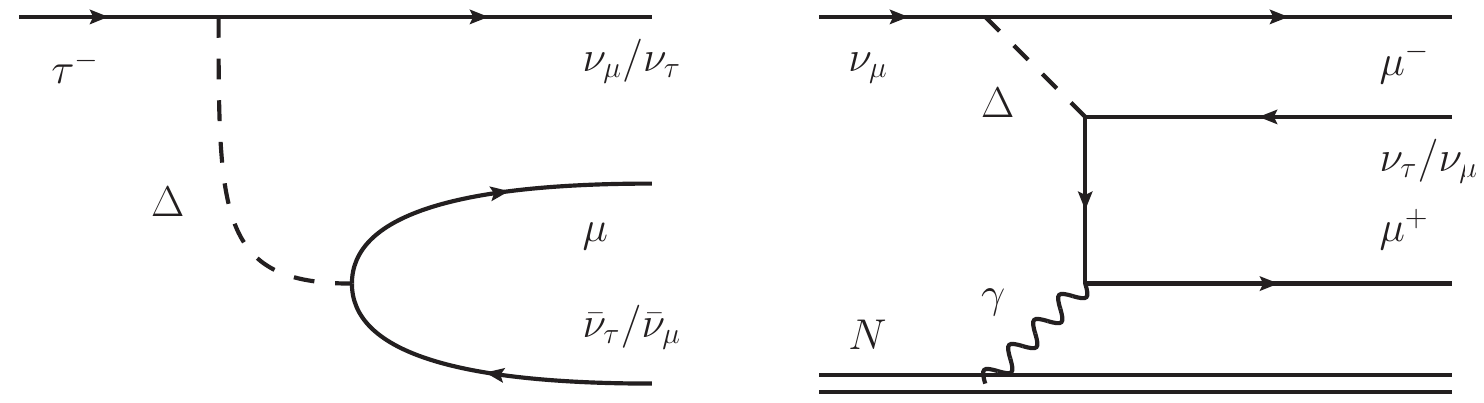}
    \end{minipage}
 \caption{The decay $\tau\to \mu\nu\bar\nu$ and neutrino trident process $\nu_\mu+N\to \nu_i+N+\mu^++\mu^-$  for singly-charged Higgs from weak triplet. 
}
 \label{tripletdecay}
\end{figure}

The above interaction in Eq.~\ref{triplet} can induce the neutrino trident $\nu_\mu+N\to \nu_{\mu,\tau} +N +\mu^+\mu^-$ at tree-level as shown in the right panel of Fig.~\ref{tripletdecay}.
The  interaction term, $\sqrt{2}\overline{\nu_\mu^c}(y_{22}(\Delta^+_{2}+\Delta^+_{3})/2-y_{23}\Delta^+_{1})\mu_L+h.c.$, contributes to  the corresponding  ratio as
\begin{eqnarray}\label{sigmaDelta^+}
   \frac{\sigma_{\Delta^+}}{  \sigma_{SM}}|_{trident}=
   \frac{\left[1+4 s^2_W- \frac{m^2_W}{g^2}\left(2\frac{y_{22}^2}{m_{\Delta^+_2}^2}+\frac{4y_{23}^2}{m_{\Delta^+_1}^2}\right)\right]^2+ 
    \left[1- \frac{m^2_W}{g^2}\left(2\frac{y_{22}^2}{m_{\Delta^+_2}^2}+\frac{4y_{23}^2}{m_{\Delta^+_1}^2}\right)\right]^2} 
   {[(1+4 s^2_W)^2 + 1]}\;,
\end{eqnarray}
here we use the degenerate condition $m_{\Delta^+_2}=m_{\Delta^+_3}$ required by the discrete symmetry.
We find that the $\Delta^+$ contribution will reduce the SM prediction, which makes $\sigma_{\Delta^+}/\sigma_{SM}<1$.
Based on the averaged value $0.95\pm0.25$ from the experimental results in Eq.~\ref{tridentexp}, we can obtain the upper limit $\left(\frac{y_{22}^2}{m_{\Delta^+_2}^2}+\frac{2y_{23}^2}{m_{\Delta^+_1}^2}\right)<2\times 10^{-5}$ GeV$^{-2}$.

Similarly, the interaction described in Eq.~\ref{triplet} also induces the decay 
$\tau \to \mu \bar{\nu} \nu$ as shown in the left panel of Fig.~\ref{tripletdecay}.
However, the process involving same-flavor neutrino final states, $\tau \to \mu \bar{\nu}_{\tau} \nu_{\tau} + \mu \bar{\nu}_{\mu} \nu_{\mu}$, will not occur because $\Delta^+_{2,3}$ cannot mediate both flavor-conserving and flavor-changing interactions simultaneously.
Therefore, the only left decay channels are 
$\tau \to \mu \bar{\nu}_{\mu} \nu_{\tau} + \mu \bar{\nu}_{\tau} \nu_{\mu}$.
Combining with the contributions from the SM $W$ boson and $Z'$ described in Eq.~\ref{SM}, we can obtain the total amplitudes as
\begin{align}
&\mathcal{M}_{total}(\tau\to \mu \nu\bar{\nu})=\left(-\frac{g^2}{4m_{W}^2}+\frac{\tilde g^2}{m_\tau^2-m_{Z'}^2}+\frac{y_{22}^2}{4m_{\Delta^+_{2}}^2}\right)\bar{\mu}\gamma^{\mu}\tau\bar{\nu}_{\tau}\gamma_{\mu}P_{L}\nu_{\mu}\\ \nonumber
 &-\left(-\frac{g^2}{4m_{W}^2}+\frac{y_{22}^2}{4m_{\Delta^+_{2}}^2}\right)\bar{\mu}\gamma^{\mu}\gamma_{5}\tau\bar{\nu}_{\tau}\gamma_{\mu}P_{L}\nu_{\mu}\\ \nonumber
 &+\left[\left(\frac{y_{22}^2}{4m_{\Delta^+_{2}}^2}+\frac{\tilde g^2}{m_\tau^2-m_{Z'}^2}-\frac{y_{23}^2}{2m_{\Delta^+_{1}}^2}\right)\bar{\mu}\gamma_{\mu}\tau
 -\left(\frac{y_{22}^2}{4m_{\Delta^+_{2}}^2}-\frac{y_{23}^2}{2m_{\Delta^+_{1}}^2}\right)\bar{\mu}\gamma_{\mu}\gamma_{5}\tau\right]\overline{\nu_{\mu}}\gamma^{\mu}P_{L}\nu_{\tau}\;.
\end{align}
Here, we find that $\Delta^+_{2,3}$ will weaken the contributions from the SM $W$ boson and  large mass $Z'$ due to the opposite sign, which is in stark contrast to the weak singlet case described in Eq.~\ref{taudecaysinglet}.
Additionally, $\Delta^+_{1}$ will further reduce the effects of $\Delta^+_{2,3}$, as shown in the last line.
In fact, the contribution in the last line is proportional to the quartic couplings, which are much smaller than those from the first two terms with quadratic couplings.
Correspondingly, the ratio $R^{\tau}$ can be obtained as
\begin{eqnarray}
     R^\tau=&\left(1-\frac{4m_W^2}{g^2}\frac{y_{22}^2}{4m_{\Delta_2^+}^2}\right)^2+\frac{4m_W^2}{g^2}\frac{\tilde g^2}{m_{Z^{\prime}}^2}\left(1-\frac{4m_W^2}{g^2}\frac{y_{22}^2}{4m_{\Delta_2^+}^2}
  \right)f\left(\frac{m_\tau^2}{m_{Z^{\prime}}^2}\right)+ \frac{\tilde g^4}{g^4}\frac{4m_W^4}{m_{Z^{\prime}}^4}g\left(\frac{m_\tau^2}{m_{Z^{\prime}}^2}\right)\;.
\end{eqnarray}
here we simplify the expression by adopting the  condition $m_{\Delta^+_2}=m_{\Delta^+_3}$.

Similar to the decay $\tau \to \mu \nu\bar\nu$ mediated by $\Delta_i^+$, $\Delta_i^{++}$ can also contribute to the decay $\tau \to 3\mu$ by converting  neutrinos into charged muons  with 
\begin{eqnarray}
    Br(\tau\to 3\mu)=\frac{m_\tau^5\tau_\tau}{3(16\pi)^3}\left(\frac{y^2_{22}}{m^2_{\Delta_2^{++}}}-\frac{y^2_{22}}{m^2_{\Delta_3^{++}}}\right)^2=0
\end{eqnarray}
Fortunately, the exchange symmetry enforces $m_{\Delta^{++}_2}=m_{\Delta^{++}_3}$, which which prohibits the aforementioned decays. A similar situation occurs in the decay $\tau\to \mu\gamma$.

\begin{figure*}[!t]
	\centering
	\begin{subfigure}[b]{0.49\textwidth}
		\centering
		\includegraphics[width=\textwidth]{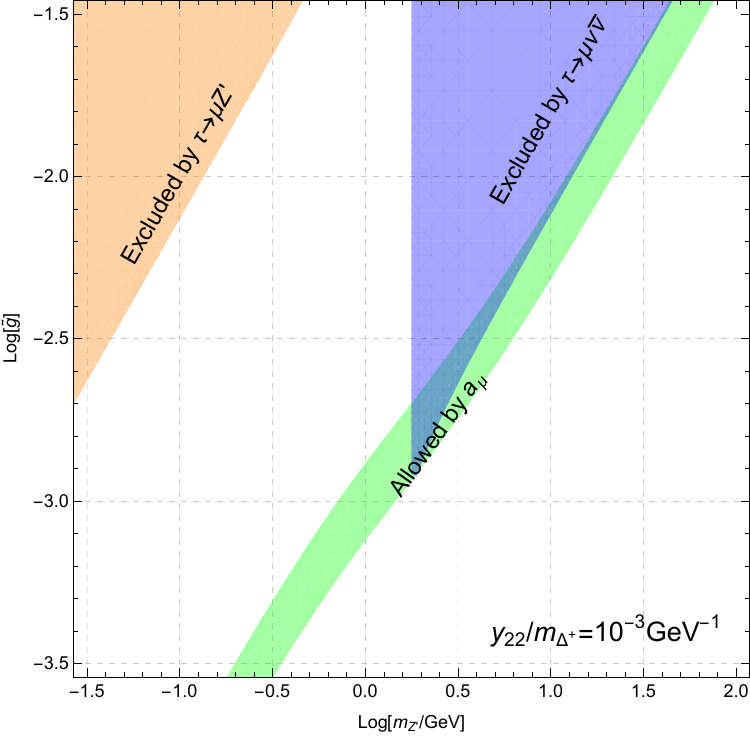}
		\caption{}
		\label{fig:triplet2}
	\end{subfigure}
	\hfill
	\begin{subfigure}[b]{0.49\textwidth}
		\centering
		\includegraphics[width=\textwidth]{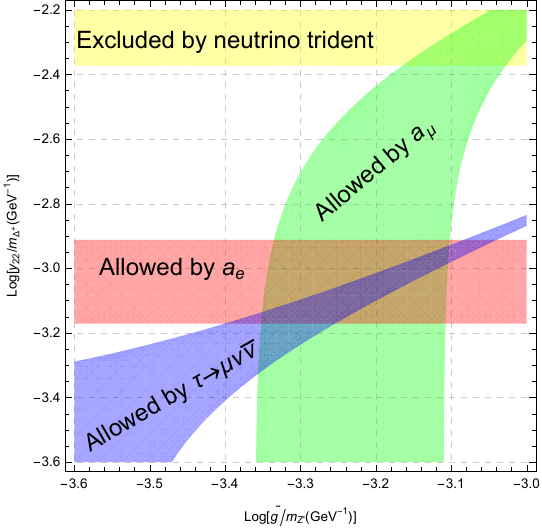}
		\caption{}
		\label{fig:triplet1}
	\end{subfigure}
	\caption{The allowed parameter regions for the maximal flavor-changing $Z'$ models with singly-charged Higgs from weak triplet (Case II). The red and green regions indicate the $(g-2)_e$ and $(g-2)_\mu$ allowed ranges within 2$\sigma$ errors, respectively. The excluded regions are shown in orange and blue: $\tau\to\mu Z'$ for small $m_{Z'}$ and $\tau\to\mu \nu\bar\nu$ for $m_{Z'} > m_\tau - m_\mu$, respectively. a) shows the $\log[m_{Z'}/\text{GeV}]-\log[\tilde g]$ plane with $y_{22}/m_{\Delta^+}=10^{-3}$GeV$^{-1}$. b) shows the $\log[\tilde g/m_{Z'}(\text{GeV}^{-1})]-\log[y_{23}/m_{\Delta^+}(\text{GeV}^{-1})]$ plane for the large $m_{Z'}$ case.}
	\label{fig:triplet}
\end{figure*}

Based on the preceding analysis, we plot the corresponding parameter regions, as illustrated in Fig.~\ref{fig:triplet}.
To simplify the analysis and highlight the viable parameter regions, we set $y_{23}=0$ to ensure the processes $\tau \to e\nu\bar\nu$ and $\mu \to e\nu\bar\nu$ unaffected because we choose non-zero $y_{11}$ for $(g-2)_e$. If we do not case about $a_e$ anomaly, alternative choices are possible, such as
$\frac{y_{22}^2}{4m_{\Delta^+_{2}}^2}=\frac{y_{23}^2}{2m_{\Delta^+_{1}}^2}$. 
The green regions represent the $(g-2)_\mu$ values allowed within 2$\sigma$ uncertainties.
In Fig.~\ref{fig:triplet2}, we fix the charged Higgs parameter at $y_{22}/m_{\Delta^+}=10^{-3}$GeV$^{-1}$.
The excluded regions for $\tau\to\mu Z'$ (for small $m_{Z'}$) and $\tau\to\mu \nu\bar\nu$ (for $m_{Z'}>m_\tau-m_\mu$) are shown in orange and blue, respectively.
We find that there are viable parameter ranges that reconcile the discrepancy between $(g-2)_\mu$ and $\tau \to \mu\nu\bar\nu$ across the entire $m_{Z'}$ range.
In Fig.~\ref{fig:triplet1}, we denote large $m_{Z'}$ values, on the order of hundreds of GeV.
The neutrino trident constraint excludes the parameter region for charged Higgs, which is highlighted in yellow.
 The allowed regions by $a_e$, $a_\mu$, and $\tau\to \mu\nu\bar\nu$, are shown in red, green, and blue, respectively.
 We choose $y_{11}/m_{\Delta_1^+}=25 y_{22}/m_{\Delta_2^+}$ to ensure the $a_e$ intersection of the intervals with $a_\mu$ and $\tau\to \mu\nu\bar\nu$.
 We find that there are common regions, specifically $y_{22}/m_{\Delta^+}=(10^{-3.15}-10^{-2.95})$ GeV$^{-1}$  and $\tilde g/m_{Z'}=(10^{-3.35}-10^{-3.12})$ GeV$^{-1}$, that can explain the $(g-2)_{e,\mu}$ anomaly while satisfying the other constraints.

  Therefore, we find that the singly-charged Higgs $\Delta^+$ from  weak triplet can reconcile the discrepancy between $(g-2)_\mu$ and $\tau\to\mu\nu\bar\nu$ in the case of large $m_{Z'}$.
  Furthermore, it can also explain the electron $(g-2)_e$ anomaly.
  This suggests that the weak triplet scalar could play a crucial role in studying the $Z'$ model.

 \subsection{ singly charged Higgs from weak doublet (Case III)}

 In this subsection, we analyze the LR operator case presented in Eq.~\ref{operator}.
 Analogous to the SM Yukawa coupling, the LR operator involving a singly charged scalar and leptons can be readily realized with a new weak doublet Higgs $H'$, which has
\begin{align}
\mathcal{L}^{LR}=g_{ab}\overline{L}_{La} H' e_{Rb}+h.c.
=g_{ab}(\overline{\nu_{La}} H'^+ e_{Rb}+ \overline{e_{La}} H'^0 e_{Rb})+h.c.\;,  
\end{align}
here $L_{L}$ means the left-handed lepton doublet, and $e_{R}$ means the right-handed charged lepton singlet. $a,b=1,2,3$ stands for the generation of leptons.

Actually,  the LR operator can also be realized at dimension-5 using only a singly-charged singlet with
\begin{align}
\mathcal{L}_{h^{\pm}}^{LR}=\frac{c_{ab}}{\Lambda}\bar{\ell}_{La}\tilde{H}e_{Rb}h^{+}+h.c.\;, \longrightarrow   \mathcal{L}_{h^{\pm}}^{LR}=\frac{c_{ab}}{\Lambda}\frac{v}{\sqrt{2}}\bar{\nu}_{La}e_{Rb}h^{+}+h.c.\;,
\end{align}
here SM Higgs $\tilde H=i\sigma^2 H^*$, and $c_{ab}$ encodes the information about new degrees of freedom with scale  $\Lambda$.
After spontaneous symmetry breaking, we can naturally obtain the required LR operator. 
Here $v=246$ GeV is SM Higgs  vacuum expectation value.

\begin{figure}[htbp!]
    \begin{minipage}[t]{0.9\linewidth}
  \centering
  \includegraphics[width=1.0\columnwidth]{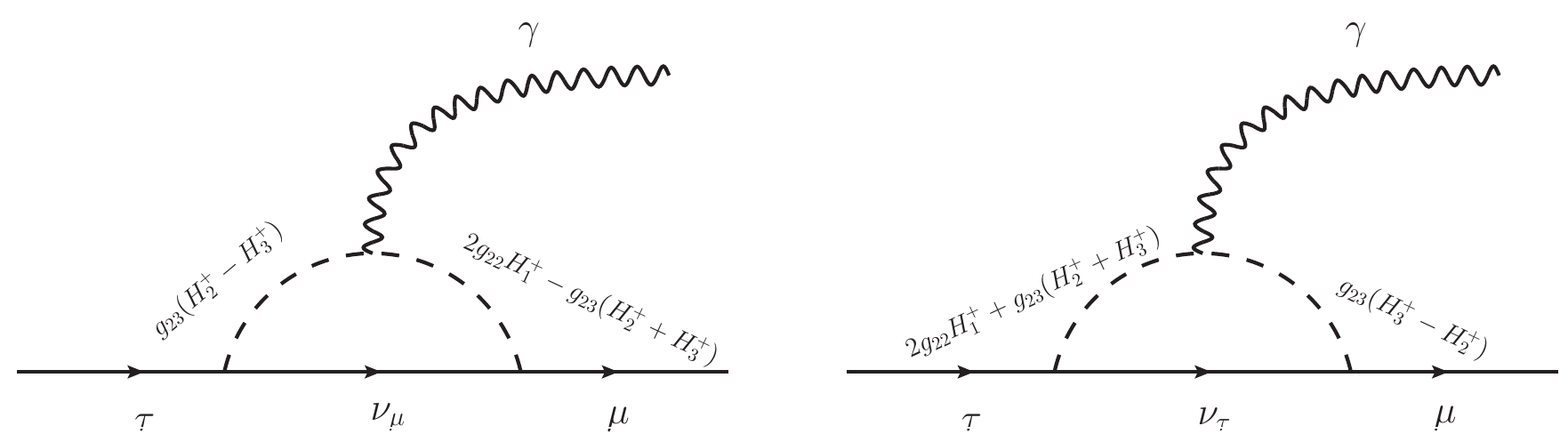}
    \end{minipage}
 \caption{ The decay $\tau \to \mu\gamma$ from the charged Higgs $H^+_{1,2,3}$ induced by weak doublet.} \label{taudecay}
\end{figure}

 Both methods outlined above can generate the required LR operator. Therefore, we will focus on the general expression, given by
\begin{align}
\mathcal{L}_{H^{\pm}}^{LR}=g_{ab}\bar{\nu}_{La}e_{Rb}H^{+}+h.c.\;,  
\end{align}
Similarly, we can introduce three charged Higgs bosons: $H^+_1(0)$, $H^+_2(2)$, and $H^+_3(-2)$, to preserve the discrete exchange symmetry.
 The numbers in the bracket stand for  the corresponding $U(1)_{L_{\mu}-L_{\tau}}$ charges.
Based on these $U(1)_{L_{\mu}-L_{\tau}}$ charges, we can construct the corresponding Yukawa coupling, given by
\begin{align}
\mathcal{L}_{LR}=&
g_{11}\bar{\nu}_{L1}e_{R1}H_{1}^{+}+g_{22}\bar{\nu}_{L2}e_{R2}H_{1}^{+}+g_{33}\bar{\nu}_{L3}e_{R3}H_{1}^{+}
+g_{23}\bar{\nu}_{L2}e_{R3}H_{2}^{+}+g_{32}\bar{\nu}_{L3}e_{R2}H_{3}^{+}+h.c.\nonumber\\
=
&\begin{pmatrix}\overline{\nu}_{L1},&\overline{\nu}_{L2},&\overline{\nu}_{L3}\end{pmatrix}\begin{pmatrix} g_{11}H_{1}^+ &0&\\0&g_{22}H_{1}^+ &g_{23}H_{2}^+\\0&g_{23}H_{3}^+ &g_{22}H_{1}^+\end{pmatrix}\begin{pmatrix}e_{R1}\\e_{R2}\\e_{R3}\end{pmatrix}+h.c. 
\end{align}
In the second equation, we use the relation of the coupling constant $g_{23}=g_{32}$ and $g_{22}=g_{33}$ from the discrete exchange symmetry. 
Furthermore, we use the diagonalized transformation in Eq.~\ref{eigen} between the charged lepton mass eigenstate and flavor eigenstate basis. Then we can obtain the couplings as
\begin{align}\label{case3}
\mathcal{L}_{LR}=\frac{1}{2}
&\begin{pmatrix}\overline{\nu}_{Le},&\overline{\nu}_{L\mu},&\overline{\nu}_{L\tau}\end{pmatrix}\begin{pmatrix} 2g_{11}H_{1}^+ &0&0\\0&2g_{22}H_{1}^+ - g_{23}H_{2+3}^{+}&g_{23}H_{2-3}^{+}\\0&-g_{23}H_{2-3}^{+} &2g_{22}H_{1}^+ + g_{23}H_{2+3}^{+}\end{pmatrix}\begin{pmatrix}e_{R}\\\mu_{R}\\\tau_{R}\end{pmatrix}+h.c.,
\end{align}
with $H_{2+3}^{+}=H_{2}^+ +H_{3}^+$ and $H_{2-3}^{+}=H_{2}^+ -H_{3}^+$.
Note that  $H_{2,3}^+$ can  mediate both the flavor-conserving and -changing interactions between the tau and muon leptons, while $H_{1}^+$ only has the flavor-conserving interactions.
And the feature with different sign  in the diagonal and off-diagonal terms for $H^+_{2,3}$ cancels out each other, such as $\tau\to \mu\gamma$ in Fig.~\ref{taudecay}. 
We find that the neutrinos running the loop diagram are the  muon $\nu_\mu$ or  tau neutrinos  $\nu_\tau$, which both involve in diagonal and off-diagonal terms.

\begin{figure}[htbp!]
  \begin{minipage}[t]{1\linewidth}
  \centering
  \includegraphics[width=1.0\columnwidth]{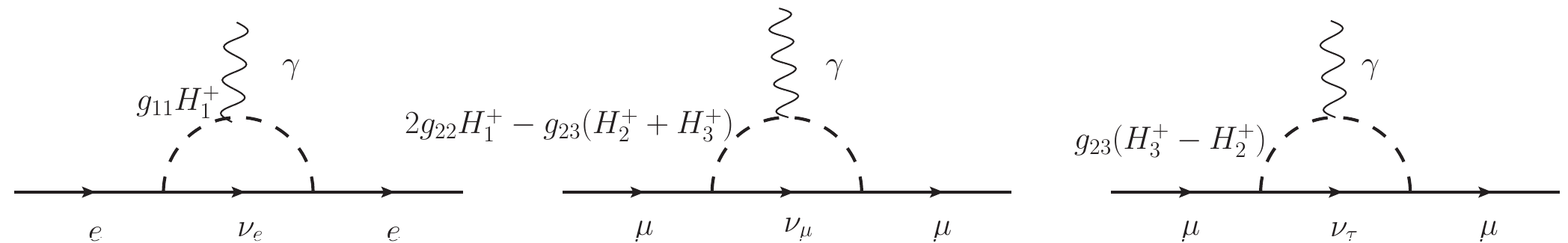}
    \end{minipage}
 \caption{ The charged  lepton $(g-2)_l$   from the charged Higgs $H^+_{1,2,3}$ induced by weak doublet.}
 \label{g2}
\end{figure}

In this case, the $H^+_{1,2,3}$ can generate the lepton  $(g-2)_{e,\mu}$  as shown in Fig.~\ref{g2}.
Based on the general expression in Eq.~\ref{chargedg2}, the corresponding contribution is
\begin{eqnarray}
 a_{e}&=&-\frac{m_{e}^2}{96\pi^2}\frac{g_{11}^2}{m_{H_1^{+}}^2}\;,\nonumber\\
  a_{\mu}&=&-\frac{m_{\mu}^2}{96\pi^2}\frac{g_{22}^2}{m_{H_1^{+}}^2}  
   -\frac{m_{\mu}^2}{96\pi^2}
   \left(\frac{g_{23}^2}{4m_{H_2^{+}}^2}
   +\frac{g_{23}^2}{4m_{H_3^{+}}^2}\right)\times 2
   = -\frac{m_{\mu}^2}{96\pi^2}\frac{g_{22}^2}{m_{H_1^{+}}^2}  
   -\frac{m_{\mu}^2}{96\pi^2}\frac{g_{23}^2}{m_{H_2^{+}}^2}\;.
\end{eqnarray}
For $a_\mu$, the factor $g_{23}^2/4$ in the first equation is from 
the coupling coefficient $-g_{23}/2$, and  the additional factor 2 is due to   the neutrino flavors $(\nu_\mu,\nu_\tau)$ running in the loop.
In the second equation for $a_\mu$, we use the degenerate condition $m_{H_2^+}=m_{H_3^+}$ required by the discrete  exchange symmetry.

\begin{figure}[htbp!]
 \begin{minipage}[t]{0.49\linewidth}
  \centering
  \includegraphics[width=0.7\linewidth]{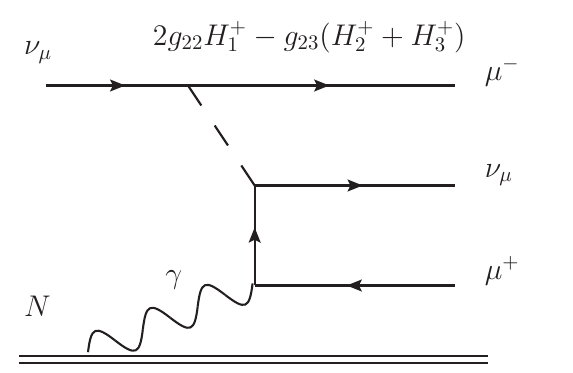}
    \end{minipage}
     \begin{minipage}[t]{0.49\linewidth}
  \centering
  \includegraphics[width=0.7\linewidth]{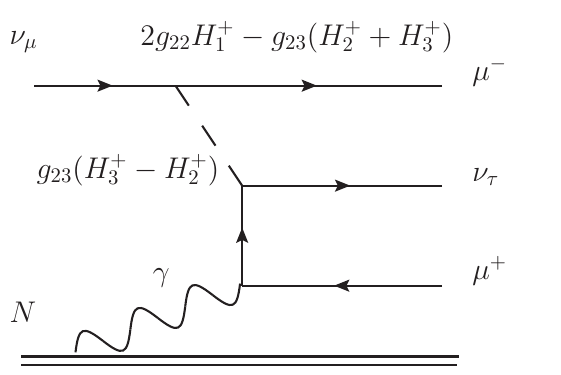}
    \end{minipage}
 \caption{ The neutrino trident production $\nu_\mu+N\to \nu_i +N+\mu^+\mu^-$ from the charged Higgs $H^+_{1,2,3}$ induced by weak doublet.} \label{NMT}
\end{figure}

The above interaction in Eq.~\ref{case3} can induce the neutrino trident $\nu_\mu+N\to \nu_i +N +\mu^+\mu^-$ at tree-level by the interaction term $\overline{\nu_\mu}(g_{22}H_1^+-g_{23}(H^+_{2}+H^+_{3})/2)\mu_L+h.c.$, with the corresponding Feynman diagram  in Fig.~\ref{NMT}.

As shown in Fig.~\ref{NMT}, the second diagram will disappear due to the opposite coupling sign  for $H_2^+$ and  $H_3^+$.
Therefore, the only left diagram is the first one with the Lagrangian as
\begin{align}
\mathcal{L}_{H}=-\left(\frac{g_{22}^2}{8m_{H_{1}^{+}}^2}+\frac{g_{23}^2}{32m_{H_{2,3}^{+}}^2}\right)\bar{\mu}\gamma_{\mu}(1+\gamma_{5})\mu\bar{v}_{\mu}\gamma^{\mu}(1-\gamma_{5})v_{\mu}\;.
\end{align}
This correspondingly results in the ratio as
\begin{eqnarray}
   \frac{\sigma_{H^+}}{  \sigma_{SM}}|_{trident}=
   \frac{\left[1+4 s^2_W- \frac{m^2_W}{g^2}\left(2\frac{g_{22}^2}{m_{H^+_1}^2}+\frac{g_{23}^2}{m_{H^+_2}^2}\right)\right]^2+ 
    \left[1+ \frac{m^2_W}{g^2}\left(2\frac{g_{22}^2}{m_{H^+_1}^2}+\frac{g_{23}^2}{m_{H^+_2}^2}\right)\right]^2} 
   {[(1+4 s^2_W)^2 + 1]}\;,
\end{eqnarray}
Compared to the case in Eq.~\ref{sigmaDelta^+} induced by weak triplet, the only difference is the positive sign in the second term.
Similarly, the averaged data $0.95\pm0.25$ from  Eq.~\ref{tridentexp} can constrain the upper limit $\left(2\frac{g_{22}^2}{m_{H^+_1}^2}+\frac{g_{23}^2}{m_{H^+_2}^2}\right)<1\times 10^{-4}$ GeV$^{-2}$.

\begin{figure}[htbp!]
        \begin{minipage}[t]{0.8\linewidth}
  \centering
  \includegraphics[width=1.0\columnwidth]{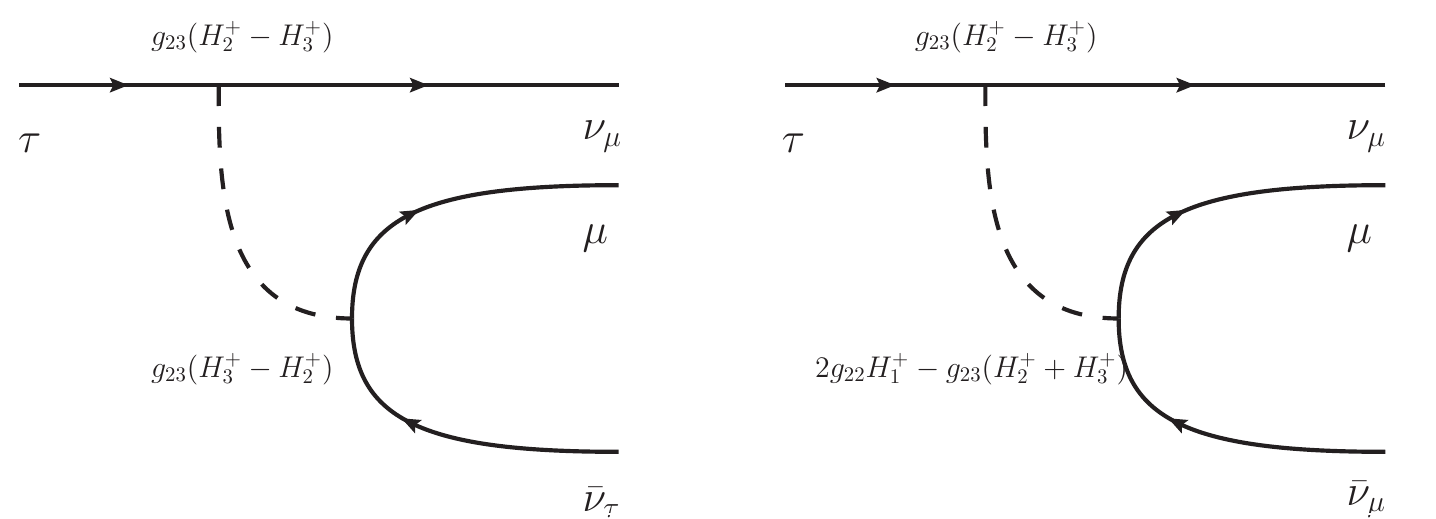}
    \end{minipage}
        \begin{minipage}[t]{0.8\linewidth}
  \centering
  \includegraphics[width=1.0\columnwidth]{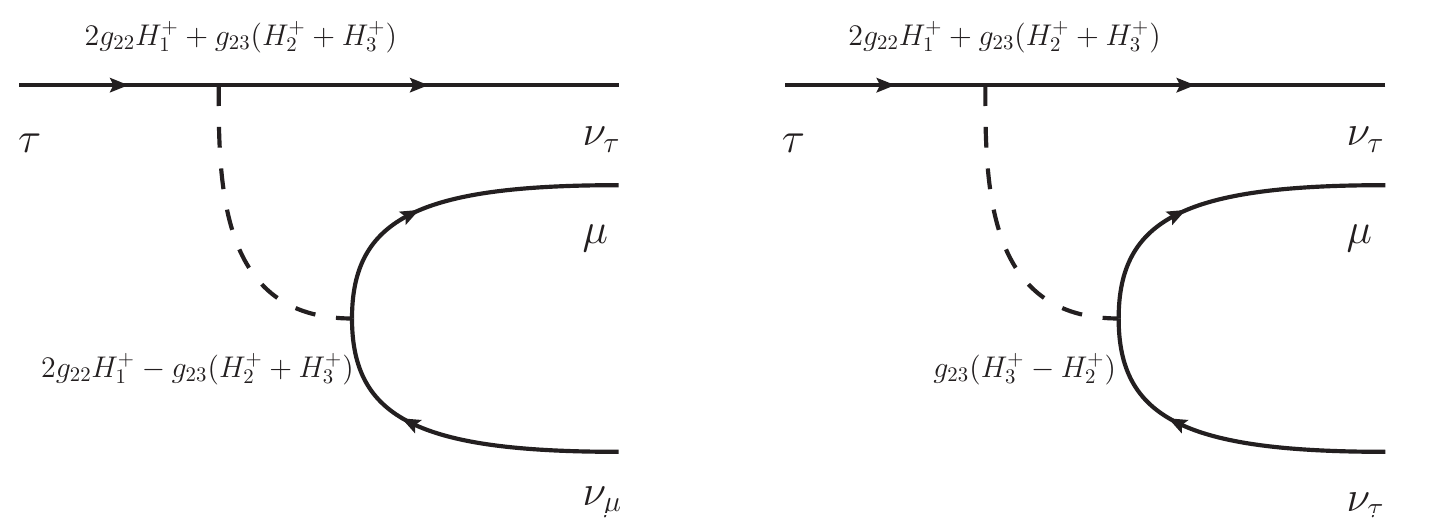}
    \end{minipage}
 \caption{ The tau decay $\tau\to \mu\nu\bar\nu$ from the charged Higgs $H^+_{1,2,3}$ induced by weak doublet.
} \label{decay width}
\end{figure}

The interaction in Eq.~\ref{case3} also induces the decay  $\tau\to\mu\bar{\nu}\nu$ as shown in Fig.~\ref{decay width}. 
However, the process  involving in the same flavor neutrino final states, $\tau\to\mu\bar{\nu}_{\tau}\nu_{\tau}+\mu\bar{\nu}_{\mu}\nu_{\mu}$,  will disappear since the  $H^+_{2,3}$ cannot mediate the flavor-conserving and -changing interactions simultaneously.
Therefore, the left decay channels are 
$\tau\to\mu\bar{\nu}_{\mu}\nu_{\tau}+\mu\bar{\nu}_{\tau}\nu_{\mu}$ as show in the left two Feynman diagrams  in Fig.~\ref{decay width}.
Then we can the interaction Lagrangian  as
\begin{align}
\mathcal{L}_{H^{+}}
=&\left(\frac{g_{23}^2}{8m_{H_{2}^{+}}^2}+ \frac{g_{23}^2}{8m_{H_{3}^{+}}^2}-\frac{g_{22}^2}{2m_{H_{1}^{+}}^2}\right)\bar{\nu}_{L\tau}\gamma_{\mu}P_{L}\nu_{\mu}\bar{\mu}_{R}\gamma^{\mu}P_{R}\tau_{R}\\ \nonumber
&+\left(\frac{g_{23}^2}{8m_{H_{2}^{+}}^2}+\frac{g_{23}^2}{8m_{H_{3}^{+}}^2}\right)\bar{\nu}_{L\mu}\gamma_{\mu}P_{L}\nu_{\tau}\bar{\mu}_{R}\gamma^{\mu}P_{R}\tau_{R}\;,
\end{align}
here we use  the Fierz transformation $\bar{u}_{1}\gamma^{\mu}(1-\gamma_{5})u_{2}\bar{u}_{3}\gamma^{\mu}(1+\gamma_{5})u_{4}=-\frac{1}{2}\bar{u}_{1}\gamma^{\mu}(1-\gamma_{5})\bar{u}_{4}\bar{u}_{3}\gamma_{\mu}(1+\gamma_{5})u_{2}$.
And we found that the charged lepton is right-handed, which is totally different from the case of  LL operators.

When incorporating with the contribution from SM $W$ boson and $Z'$ in Eq.~\ref{SM}, we can obtain the total decay amplitudes as 
\begin{align}
M(\tau\to \mu\nu\bar\nu)&=\left(-\frac{g^2}{4m_{W}^2}+\frac{\tilde g^2}{m_\tau^2-m_{Z'}^2}+\frac{g_{23}^2}{8m_{H_{2}^{+}}^2}-\frac{g_{22}^2}{4m_{H_{1}^{+}}^2}\right)\bar{\nu}_{\tau}\gamma_{\mu}P_{L}\nu_{\mu}\bar{\mu}\gamma^{\mu}\tau\\ \nonumber
&+\left(\frac{g^2}{4m_{W}^2}+\frac{ g_{23}^2}{8m_{H^+_2}^2}-\frac{g_{22}^2}{4m_{H_{1}^{+}}^2}\right)\bar{\nu}_{L\tau}\gamma_{\mu}P_{L}\nu_{\mu}\bar{\mu}\gamma^{\mu}\gamma_{5}\tau\\ \nonumber
&+\left[\left(\frac{g_{23}^2}{8m_{H_{2}^{+}}^2}+\frac{\tilde g^2}{m_\tau^2-m_{Z'}^2}\right)\bar{\mu}\gamma_{\mu}\tau+\frac{g_{23}^2}{8m_{H_{2}^{+}}^2}\bar{\mu}\gamma_{\mu}\gamma_{5}\tau\right]\bar{\nu}_{L\tau}\gamma_{\mu}P_{L}\nu_{\mu}\;.
\end{align}
here we use the degenerate condition $m_{H^+_2}=m_{H^+_3}$ required by the discrete exchange symmetry. We found that the first two lines about $H^+$ are canceled  out at the order $O(\tilde g^2/m_{H^+}^2)$, and the $H^+$ contribution only appears in $\tilde g^4/m_{H^+}^4$.
Correspondingly, the ratio can be characterized as  
 \begin{align}
 R^{\tau}&= 1+\frac{\tilde g^2}{g^2}\frac{4m_W^2}{m_{Z^{\prime}}^2}f\left(\frac{m_\tau^2}{m_{Z^{\prime}}^2}\right)+\frac{\tilde g^4}{g^4}\frac{4m_W^4}{m_{Z^{\prime}}^4}g\left(\frac{m_\tau^2}{m_{Z^{\prime}}^2}\right)\notag\\
  &+\frac{4m_W^4}{g^4}\frac{\tilde g^2}{m_{Z^{\prime}}^2}\left(
 \frac{ g_{23}^2}{m_{H^+_2}^2}-\frac{g_{22}^2}{m_{H_{1}^{+}}^2}\right)f\left(\frac{m_\tau^2}{m_{Z^{\prime}}^2}\right)
+\frac{4m_W^4}{g^4}\left[\left(\frac{ g_{23}^2}{2m_{H^+_2}^2}-\frac{g_{22}^2}{m_{H_{1}^{+}}^2}\right)^2+\frac{ g_{23}^4}{4m_{H^+_2}^4}\right]\;,
 \end{align}
 here 
 we found this expression keeps the approximate form with Eq.~\ref{Z'tau} for the maximal flavor-changing $Z'$ model since the term $g^4/m_{H^+}^4$ is much suppressed.
Thereby $H^+$ induced by weak doublet keeps the same situation with maximal $Z'$ model approximately.

\begin{figure*}[!t]
	\centering
	\begin{subfigure}[b]{0.49\textwidth}
		\centering
		\includegraphics[width=\textwidth]{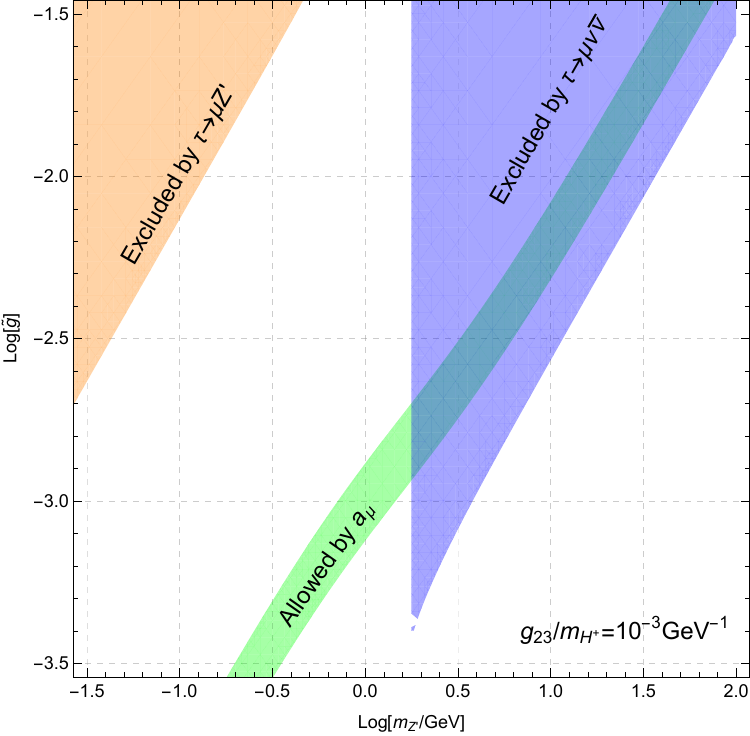}
		\caption{}
		\label{fig:doublet2}
	\end{subfigure}
	\hfill
	\begin{subfigure}[b]{0.49\textwidth}
		\centering
		\includegraphics[width=\textwidth]{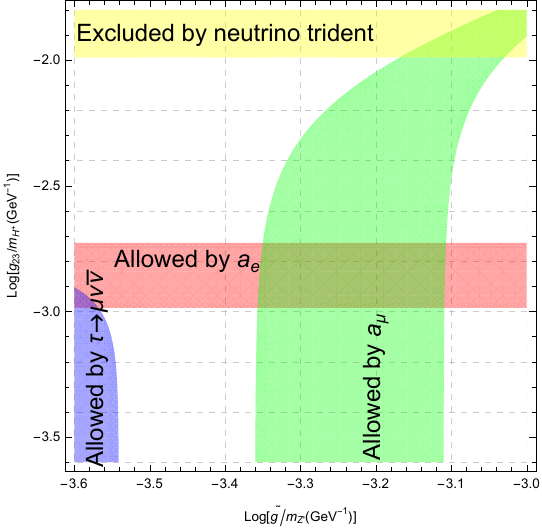}
		\caption{}
		\label{fig:doublet1}
	\end{subfigure}
	\caption{The allowed parameter regions for the maximal flavor-changing $Z'$ models with singly-charged Higgs from weak doublet (Case III). The red and green regions indicate the $(g-2)_e$ and $(g-2)_\mu$ allowed ranges within 2$\sigma$ errors, respectively. a) shows the $\log[m_{Z'}/\text{GeV}]-\log[\tilde g]$ plane with $g_{23}/m_{\Delta^+}=10^{-3}$GeV$^{-1}$. The excluded regions by $\tau\to\mu Z'$ for small $m_{Z'}$ region and $\tau\to\mu \nu\bar\nu$ for $m_{Z'}>m_\tau-m_\mu$ are shown in orange and blue, respectively. b) shows the $\log[\tilde g/m_{Z'}(\text{GeV}^{-1})]-\log[g_{23}/m_{\Delta^+}(\text{GeV}^{-1})]$ plane for the large $m_{Z'}$ case. The excluded region by the neutrino trident process is shown in yellow.}
	\label{fig:doublet}
\end{figure*}

Based on the above  analysis, we plot the corresponding parameter regions as shown in Fig.~\ref{fig:doublet}. 
For simplifying the analysis to show the viable parameter regions, we choose $g_{22}=0$ to avoid affecting  $\tau \to e\nu\bar\nu$ and $\mu \to e\nu\bar\nu$ as a non-zero $g_{11}$
is selected to explain $(g-2)_e$. If $a_e$
  is not a concern, we can make other choice, such as
$\frac{g_{23}^2}{m_{H^+_{2}}^2}=\frac{g_{22}^2}{m_{H^+_{1}}^2}$. 
 The  green regions mean the $(g-2)_\mu$ allowed ranges   within 2$\sigma$ errors. 
 In Fig.~\ref{fig:doublet2}, we fix the charged Higgs parameters $g_{23}/m_{H^+}=10^{-3}$ GeV$^{-1}$. 
 The excluded regions  by $\tau \to \mu Z'$ (for small $m_{Z'}$) and $\tau \to \mu \nu \bar{\nu}$ (for $m_{Z'} > m_\tau - m_\mu$) are shown in orange and blue, respectively.
 We find that, for large $m_{Z'}$ on the GeV scale, the $(g-2)_\mu$ allowed region is completely excluded by $\tau \to \mu \nu \bar{\nu}$.
 Moreover, the figures approximately match the regions observed in the maximal flavor-changing $Z'$ model, as shown in Fig.~\ref{Z'region}.
 In Fig.~\ref{fig:doublet1}, we illustrate large $m_{Z'}$ values on the order of hundreds of GeV.
 The neutrino trident constraint excludes the parameter region for the charged Higgs, highlighted in yellow.
 The allowed regions based on $a_e$, $a_\mu$, and $\tau \to \mu \nu \bar{\nu}$ are shown in red, green, and blue, respectively.
Similarly,  here we choose $y_{11}/m_{\Delta_1^+}=40 y_{22}/m_{\Delta_2^+}$ to ensure the $a_e$ intersection interval  with $a_\mu$ or $\tau\to \mu\nu\bar\nu$.  
We find that there are no viable common regions that can explain the $(g-2)_{e,\mu}$ anomaly while satisfying the experimental constraints from $\tau \to \mu \nu \bar{\nu}$.

 Therefore, we find that  singly-charged Higgs boson $H^+$ from  weak doublet cannot reconcile the discrepancy between $(g-2)_{\mu}$ and $\tau \to \mu \nu \bar{\nu}$ in the case of large $m_{Z'}$.
 Furthermore, it essentially maintains the same situation as the maximal flavor-changing $Z'$ model, since the contribution from the charged Higgs is negligible in terms of quartic couplings.

\section{Summary}
\label{sec:conc}

In this paper, we investigate the $U(1)_{L_\mu-L_\tau}$ $Z'$  model with singly-charged scalars and analyze the relevant phenomenology to determine the viable parameter regions.
The exchange of a $U(1)_{L_\mu-L_\tau}$ $Z'$ gauge boson is a favored mechanism for addressing the muon $(g-2)_\mu$ anomaly.
The simplest  $U(1)_{L_\mu-L_\tau}$ model features a flavor-conserving $Z'$ interaction given by $(\bar \mu \gamma^\mu \mu- \bar \tau \gamma^\mu \tau + \bar \nu^\mu_L \gamma^\mu \nu^\mu_L - \bar \nu^\tau_L \gamma^\mu \nu^\tau_L)Z'_\mu$.
The diagonal $Z'$ interactions are stringently constrained by muon neutrino trident and NA64$\mu$ experiments, which impose an upper bound on the $Z'$ mass of approximately $m_{Z'} < 40$ MeV.
The kinetic mixing between the hypercharge $U(1)_{Y}$ and $U(1)_{L_\mu-L_{\tau}}$ cannot significantly relax these constraints.
In fact, the bounds could be relaxed by introducing mixing between $\mu$ and $\tau$.
However, this mixing would induce lepton flavor-violating (LFV) processes, making it impossible to resolve the $(g-2)_\mu$ anomaly.
To circumvent these constraints, we propose converting the diagonal $Z'$ interaction into an off-diagonal one: 
 $(\bar \mu \gamma^\mu \tau+ \bar \tau \gamma^\mu \mu + \bar \nu^\mu_L \gamma^\mu \nu^\tau_L + \bar \nu^\tau_L \gamma^\mu \nu^\mu_L)Z'_\mu$.
This off-diagonal $Z'$ interaction is referred to as the maximal flavor-violating $U(1)_{L_\mu-L_\tau}$ model.
From the perspective of UV completion, the maximal flavor-violating $U(1)_{L_\mu-L_\tau}$ model can be constructed by introducing three Higgs doublets with $U(1)_{L_\mu-L_\tau}$ charges of $(0,2,-2)$ and imposing the discrete exchange symmetries $2 \leftrightarrow 3$ and $Z' \to -Z'$.
The exchange symmetries also forbid kinetic mixing effects.
The corresponding decay channels are $Z'\to  \nu_\mu \nu_\tau$ and $Z'\to  \mu \tau$, which depend on the $Z'$ mass.

The maximal flavor-violating $U(1)_{L_\mu-L_\tau}$ model induces $\tau$ lepton decays, including the family
lepton-number preserving three-body decay $\tau\to \mu \bar \nu_\mu \nu_\tau$ and the family lepton number violating decay $\tau\to \mu \bar \nu_\tau \nu_\mu$, due to indistinguishable neutrino flavors.
Current experimental data on $\tau \to \mu \nu \bar{\nu}$ exclude the $Z'$ parameter regions allowed by the muon $(g-2)_{\mu}$ for $m_{Z'} > m_\tau - m_\mu$, as shown in Fig.~\ref{Z'region}.
In order to reconcile the discrepancy between the muon $(g-2)_\mu$ and $\tau\to\mu\nu\bar\nu$, we attempt to introduce new interactions to circumvent the stringent $\tau \to \mu \nu \bar{\nu}$ constraints.
Similar to the SM charged and neutral current interactions, we introduce a singly-charged Higgs interaction that can mediate the decay process $\tau \to \mu \nu \bar{\nu}$.
To ensure discrete exchange symmetry, we introduce three singly-charged scalars with different $U(1)_{L_\mu-L_\tau}$ charges of $(0, 2, -2)$.

In general, introducing  singly-charged scalar can lead to three distinct scenarios: a weak singlet scalar (Case I), a triplet scalar (Case II), and a doublet scalar (Case III).
We analyze these three singly-charged scalar models and their corresponding phenomenology in detail.
 Note that singly-charged scalars contribute negatively to lepton $(g-2)_l$, which is in contrast to the positive contributions from the $Z'$ boson.
 This indicates that the $Z'$ boson and singly-charged Higgs bosons must compete with each other to achieve the required positive experimental results for $a_\mu^{\text{exp}}$.
 Our findings for these three cases are presented below:
\begin{itemize}
  \item For Case I,  the weak singlet generates interactions of the singly-charged Higgs $h^+$ (LL operator) with leptons through anti-symmetric Yukawa couplings.
  The anti-symmetric nature ensures that $h^+$ mediates only flavor-changing interactions, which can contribute to both the  $(g-2)_\mu$ anomaly and $\tau\to\mu \nu_\mu \bar \nu_\tau$.
  Analysis of the relevant experimental data reveals that the singly-charged Higgs $h^+$ from the weak singlet exacerbates the discrepancy between $(g-2)_\mu$ and $\tau \to \mu \nu \bar{\nu}$.
  The $Z'$ parameter regions with $m_{Z'} > m_\tau - m_\mu$ are further excluded regardless of the singly-charged Higgs parameters, as shown in Fig.~\ref{fig:singlet}.
  This restricts the allowed $Z'$ parameter space to small $m_{Z'}$ ranges, around the MeV scale.

  \item  For Case II, the weak triplet generates interactions of the singly-charged Higgs $\Delta^+$ (LL operator) with leptons, involving both flavor-conserving and flavor-changing interactions. These can lead to phenomena such as neutrino trident production, lepton $(g-2)_l$, and $\tau \to \mu \nu \bar{\nu}$.
  We found that there are viable parameter regions across the entire $m_{Z'}$ range that can explain the lepton $(g-2)_{e,\;\mu}$ anomaly while satisfying other experimental constraints from neutrino trident production and $\tau \to \mu \nu \bar{\nu}$, as shown in Fig.~\ref{fig:triplet}.
  This indicates that singly-charged scalars induced by the triplet can reconcile the discrepancy between $(g-2)_\mu$ and $\tau \to \mu \nu \bar{\nu}$ in the large $m_{Z'}$ regime, around GeV scales.
   This implies that the triplet scalar could play an essential role in studying $U(1)$ $Z'$ models.
  \item  For Case III, the weak doublet generates interactions of the singly-charged Higgs $H^+$ with leptons in terms of LR operator. 
  The LR operator can induce both flavor-conserving and flavor-changing Yukawa couplings.
  The opposite signs in the diagonal and off-diagonal entries for $H_{2,3}^+$ cancel out processes involving both flavor-conserving and flavor-changing interactions, such as $\tau\to \mu\gamma$ and $\nu_\mu+N\to \nu_\tau+N+\mu^++\mu^- $. 
  Other physical processes, such as  $(g-2)_l$, $\nu_\mu+N\to \nu_\mu+N+\mu^+\mu^-$ and $\tau\to\mu \nu_\mu \bar \nu_\tau(\nu_\tau \bar \nu_\mu)$, can also be mediated.
  The bounds on singly-charged Higgs from MNT are weaker compared to Case II because they result in the interaction with right-handed muons.
  The most stringent constraints from $\tau \to \mu \nu \bar{\nu}$ essentially yield a situation similar to that of maximal flavor-changing $Z'$ models, as the contribution from singly-charged Higgs arises from quartic couplings.
  We found that the lepton $(g-2)_l$ allowed regions have already been excluded by $\tau \to \mu \nu \bar{\nu}$, as shown in Fig.~\ref{fig:doublet}.
  This restricts the allowed $Z'$ parameter space to small $m_{Z'}$ ranges, around the MeV scale.
\end{itemize}

Our study shows that singly-charged scalars induced by weak singlets and doublets cannot efficiently reconcile the discrepancy between the muon $(g-2)_{\mu}$ and $\tau \to \mu \nu \bar{\nu}$ in the large $m_{Z'}$ case.
In contrast, singly-charged scalars induced by weak triplets can simultaneously explain the lepton $(g-2)_{e,\mu}$ anomaly while satisfying other experimental constraints.
Our findings indicate that singly-charged scalars could play an essential role in studying $U(1)_{L_\mu-L_\tau}$ $Z'$ models, especially triplet scalars.
This insight may be beneficial for future investigations into the $Z'$ model.

\section*{Acknowledgements}
We thank Prof. Xiao-Gang He for fruitful discussions and  Julian Heeck  for useful suggestions and comments.
The work of Fei Huang is supported by Natural Science Foundation of Shandong Province under Grant No.ZR2024QA138 and Jin Sun is supported by IBS under the project code, IBS-R018-D1.

\appendix
\section{ Gauge boson mass }\label{apA}

In the Appendix, we show the  description of scalar potential and provide a full parametrization of  all involved bosons masses, including electroweak gauge bosons, Higgs boson, Z' boson, scalar S,
$\Delta^0$, $\Delta^+$, $\Delta^{++}$.

In Case II of our model, we introduce the following scalars: three Higgs doublets $H_{1}(0),H_{2}(2),H_{3}(-2)$, three  triplet $\Delta_{1}(0),\Delta_{2}(2),\Delta_{3}(-2)$, and one singlet S(1). The numbers in the bracket mean the corresponding $U(1)_{L_\mu-L\tau}$ charges.
The symmetry breaking modes can be analyzed.
 Due to the new U(1)
charges in $H_{2,3}, \Delta_{2,3}$ and S, they will both contribute to $U(1)_{L_\mu-L_\tau}$
symmetry breaking while $H_{1,2,3}$ and $\Delta_{1,2,3}$ broken the electroweak symmetry.

The relevant Lagrangian is
\begin{eqnarray}
    \mathcal{L}_{II}=(D^\mu H_i)^\dagger (D_\mu H_i) + (D^\mu \Delta_i)^\dagger (D_\mu \Delta_i)
    + (D^\mu S)^\dagger (D_\mu S)-V(H_i,\Delta_i,S)\;.
\end{eqnarray}
Here $V(H_i,\Delta_i,S)$ means the scalar potential.
The covariant derivative is defined as
\begin{eqnarray}
&&D_{\mu}H_{i}=(\partial_{\mu}-ig_{2}W_{\mu}^{a}\frac{\sigma^{a}}{2}-ig_{1}B_{\mu}Y-i\tilde{g}Z'_{\mu}X_i)H_{i}\;,\nonumber \\
&&D_{\mu}\Delta_i=\partial_{\mu}\Delta_i-i\frac{g_{2}}{2}[W_{\mu}^{a}\sigma^{a},\Delta_i]-ig_{1}YB_{\mu}\Delta_i-i\tilde{g}Z'_{\mu}X_i\Delta_i\;,\nonumber\\
&&D_{\mu}S=(\partial_{\mu}-i\tilde{g}Z'_{\mu}X_S)S\;,
\end{eqnarray}
here $\sigma^a$ means the Pauli matrix, Y(X) represent hypercharge(new U(1) charges) of the corresponding scalars. And $g_2,g_1,\tilde g$ mean the corresponding coupling constant of gauge group $SU(2)_L,U(1)_Y,U(1)_{L_\mu-L_\tau}$.

After symmetry breaking, these scalars  obtain nonzero vacuum expectation values(vev), $<H_i>=v_i/\sqrt{2}$, $<\Delta_i>=\tilde v_i/\sqrt{2}$, $<S>=v_s/\sqrt{2}$. The above kinetic terms will contribute to the gauge boson masses. Using the relation $\tilde Y_\mu = c_W \tilde A_\mu -s_W \tilde Z_\mu$ and  $W^3_\mu =  s_W \tilde A_\mu +  c_W \tilde Z_\mu$, we can obtain the mass matrix as 
\begin{eqnarray}\label{zmixing}
  \mathcal{L}_m = & -& \frac{1}{2}(Z,Z')\left(\begin{matrix}g_{Z}^2
   (\frac{v_{1}^2+v_{2}^2+v_{3}^2}{4}+\tilde{v}_{1}^2+\tilde{v}_{2}^2+\tilde{v}_{3}^2)& g_{Z}\tilde{g}((v_{3}^2-v_{2}^2)+2(\tilde{v}_{3}^2-\tilde{v}_{2}^2))\\g_{Z}\tilde{g}((v_{3}^2-v_{2}^2)+2(\tilde{v}_{3}^2-\tilde{v}_{2}^2))&\tilde{g}^2(4(v_{3}^2+v_{2}^2)+4(\tilde v_{3}^2+\tilde v_{2}^2)+v_{s}^2)\end{matrix}\right)
   \Big(\begin{matrix}Z\\ Z'\end{matrix}\Big)\nonumber\\
 & - &\frac{1}{4}g_{2}^2(v_{1}^2+v_{2}^2+v_{3}^2+2(\tilde{v}_{1}^2+\tilde{v}_{2}^2+\tilde{v}_{3}^2))W_\mu^+ W^{\mu-}\;.
\end{eqnarray}
Here we define $g_Z=\sqrt{g_1^2+g_2^2}$.
Note that $H_i$ and $\Delta_i$ have both $U(1)_Y$ and $U(1)_{L_\mu-L_\tau}$
charges, there is in principle mixing between Z and Z' bosons as shown in Eq.~(\ref{zmixing}). Due to the discrete symmetry imposed, $v_2=v_3(\tilde v_2=\tilde v_3)$, the mixing will disappear.
Therefore, we can obtain the massed of gauge bosons as
\begin{eqnarray}
  &&  m_W^2=\frac{1}{4}g_{2}^2(v_{1}^2+2v_{2}^2+2(\tilde{v}_{1}^2+2\tilde{v}_{2}^2))=\frac{1}{4}g_{2}^2 v^2_{SM},\nonumber\\
  && m_Z^2=g_{Z}^2
   \left(\frac{v_{1}^2+2v_{2}^2}{4}+\tilde{v}_{1}^2+2\tilde{v}_{2}^2\right),\nonumber\\
  && m_{Z'}^2=\tilde{g}^2(8v_{2}^2+8\tilde v_{2}^2)+v_{s}^2)\;,
\end{eqnarray}
where $v_{1}^2+2v_{2}^2+2(\tilde{v}_{1}^2+2\tilde{v}_{2}^2)=v_{SM}=246$GeV.

Furthermore, we can obtain the electroweak $\rho$ parameter as 
\begin{eqnarray}
    \rho=1-\frac{2(\tilde{v}_{1}^2+\tilde{v}_{2}^2+\tilde{v}_{3}^2)}{v_{1}^2+v_{2}^2+v_{3}^2+4(\tilde{v}_{1}^2+\tilde{v}_{2}^2+\tilde{v}_{3}^2)}\;.
\end{eqnarray}
Current bounds will require small $\tilde v_i< O$(GeV).

\section{Scalar potential}\label{apB}

For the above scalars $H_{1}(0)$, $H_{2}(2)$, $H_{3}(-2)$, $S(1)$,  $\Delta_{1}(0)$, $\Delta_{2}(2)$, $\Delta_{3}(-2)$, the most general potential under the exchange symmetry $2\leftrightarrow 3$ can be expressed as
\begin{align}
V(H_i,\Delta_i,S)&=\lambda_{1}\left(H_{1}^{+}H_{1}-\frac{v_{1}^2}{2}\right)^2+\lambda_{2}(H_{2}^{+}H_{2}+H_{3}^{+}H_{3}-v_{2}^2)^2+\lambda_{3}(H_{2}^{+}H_{2}-H_{3}^{+}H_{3})^2\nonumber \\
&+\lambda_{4}\left(H_{1}^{+}H_{1}+H_{2}^{+}H_{2}+H_{3}^{+}H_{3}-\frac{v_{1}^2}{2}-v_{2}^2\right)^2+\lambda_{5}(H_{2}^{+}H_{2}H_{3}^{+}H_{3}-H_{2}^{+}H_{3}H_{3}^{+}H_{2})\nonumber \\
&+\lambda_{6}[(H_{1}^{+}H_{1})(H_{2}^{+}H_{2}+H_{3}^{+}H_{3})-H_{1}^{+}H_{2}H_{2}^{+}H_{1}-H_{1}^{+}H_{3}H_{3}^{+}H_{1}]\nonumber \\
&+\lambda_{7}[(H_{1}^{+}H_{1})(H_{2}^{+}H_{2}+H_{3}^{+}H_{3})-H_{1}^{+}H_{2}H_{1}^{+}H_{3}-H_{2}^{+}H_{1}H_{3}^{+}H_{1}]\nonumber \\
&+\lambda_{ij}(H_{i}^{+}H_{i})Tr[\Delta^{+}_{j}\Delta_{j}]+\tilde{\lambda}_{ij}H_{i}^{+}\Delta_{j}\Delta_{j}^{+}H_{i}+
(\mu_{i}H_{i}^{T}i\tau_{2}\Delta^+_{i}H_{1}+\tilde{\mu}_{i}H_{1}^{T}i\tau_{2}\Delta^+_{i}H_{i}+h.c.)
\nonumber \\
&+\lambda_{ij}^{\prime}(H_{i}^{+}H_{j})Tr[\Delta^{+}_{j}\Delta_{i}]+\tilde{\lambda}_{ij}^{\prime}H_{i}^{+}\Delta_{i}\Delta_{j}^{+}H_{j}+\lambda_{1}^{\prime}[(H_{1}^{+}H_{2})Tr[\Delta^{+}_{1}\Delta_{3}]+H_{1}^{+}H_{3}Tr[\Delta^{+}_{1}\Delta_{2}]+h.c.]
\nonumber \\
&+k_{ij}Tr[\Delta^{+}_{i}\Delta_{i}]Tr[\Delta^{+}_{j}\Delta_{j}]+\tilde{k}_{ij}Tr[\Delta^{+}_{i}\Delta_{j}]Tr[\Delta^{+}_{j}\Delta_{i}]\nonumber \\
&+\tilde{k}_{ij}Tr[\Delta^{+}_{i}\Delta_{i}\Delta^{+}_{j}\Delta_{j}]+\tilde{k}_{ij}^{\prime}Tr[\Delta^{+}_{i}\Delta_{j}\Delta^{+}_{j}\Delta_{i}]+M_{\Delta_{i}}^2Tr[\Delta_{i}^{+}\Delta_{i}]+\frac{1}{2}\mu_s^2 |S|^2+\lambda_s |S|^4
\;.
\end{align}
Actually, the above general  potential is complicated, so that we attempt to simplify the analysis. Here we assume that the mixing between $H_{i}$ and  $\Delta_{i}$ can be ignored with all $\lambda_{ij}=0$. Additionally, we further assume  no any intersection for different triplets, which means $i=j$. Therefore we can obtain the following potential as
\begin{align}\label{potential}
V(H_i,\Delta_i,S)&=\lambda_{1}\left(H_{1}^{+}H_{1}-\frac{v_{1}^2}{2}\right)^2+\lambda_{2}(H_{2}^{+}H_{2}+H_{3}^{+}H_{3}-v_{2}^2)^2+\lambda_{3}(H_{2}^{+}H_{2}-H_{3}^{+}H_{3})^2\nonumber \\
&+\lambda_{4}\left(H_{1}^{+}H_{1}+H_{2}^{+}H_{2}+H_{3}^{+}H_{3}-\frac{v_{1}^2}{2}-v_{2}^2\right)^2+\lambda_{5}(H_{2}^{+}H_{2}H_{3}^{+}H_{3}-H_{2}^{+}H_{3}H_{3}^{+}H_{2})\nonumber \\
&+\lambda_{6}[(H_{1}^{+}H_{1})(H_{2}^{+}H_{2}+H_{3}^{+}H_{3})-H_{1}^{+}H_{2}H_{2}^{+}H_{1}-H_{1}^{+}H_{3}H_{3}^{+}H_{1}]\nonumber \\
&+\lambda_{7}[(H_{1}^{+}H_{1})(H_{2}^{+}H_{2}+H_{3}^{+}H_{3})-
H_{1}^{+}H_{2}H_{1}^{+}H_{3}-H_{2}^{+}H_{1}H_{3}^{+}H_{1}]\nonumber \\
&+\frac{1}{2}\mu_{s}^{2}|S|^4+\lambda_{s}|S|^4+\tilde{M}_{\Delta_{i}}^2Tr[\Delta_{i}^+\Delta_{i}]+\tilde{\lambda}_{2i}(Tr[\Delta_{i}^+\Delta_{i}])^2+\tilde{\lambda}_{3i}Tr[\Delta_{i}^+\Delta_{i}\Delta_{i}^+\Delta_{i}]\;.
\end{align}
Firstly, we analyze the doublet scalar $\lambda_{1-7}$ to identity the SM Higgs bosons. The neutral components of doublet scalar fields are
\begin{align}
H_{1}=\frac{v_{1}+h_{1}+i\eta_{1}}{\sqrt{2}}\;,\quad~H_{2,3}=\frac{v_{2}+h_{2,3}+i\eta_{2,3}}{\sqrt{2}}\;,
\end{align}
Due to the imposed  discrete symmetry, it is convenient to define new eigenstate basis as 
\begin{eqnarray}
   h_{\pm}=\frac{h_{2}\pm h_{3}}{\sqrt{2}}\;,\quad \eta_{\pm}=\frac{\eta_{2}+\eta_{3}}{\sqrt{2}}\;, 
\end{eqnarray}
here the subscripts with plus and minus denote exchange-even and -odd, respectively. Note that the odd field $h_-$ have no mix with even case $h_1,h_+$.
It results in  $m_{h^{-}}^2=4\lambda_{3}v_{2}^2+\lambda_{7}v_{1}^2$.

For even field $h_1,h_+$, the mass matrix in the basis $h_1,h_+$ is 
\begin{align}
m^2_{h_{1},h^{+}}=\Big(\begin{matrix}2v_{1}^2(\lambda_{1}+\lambda_{4})&2\sqrt{2}v_{1}v_{2}\lambda_{4}\\2\sqrt{2}v_{1}v_{2}\lambda_{4}&4v_{2}^2(\lambda_{2}+\lambda_{4})\end{matrix}\Big)\;.
\end{align}
We found that the mass matrix is not diagonal. We should conduct the diagonalization. 
The transformation between mass eigenstate and flavor eigenstate basis is given by
\begin{align}
\left(\begin{matrix}h^{\prime}\\h_{+}^{\prime}\end{matrix}\right)=\left(\begin{matrix}\cos\theta&\sin\theta \\ -\sin\theta&\cos\theta\end{matrix}\right)\left(\begin{matrix}h_{1}\\h_{+}\end{matrix}\right)\;,\quad
\tan2\theta=\frac{2\sqrt{2}v_{1}v_{2}\lambda_{4}}{(\lambda_{1}+\lambda_{4})v_{1}^2-2(\lambda_{2}+\lambda_{4})v_{2}^2}\;.
\end{align}
Therefore the masses of SM Higgs and another exchange-even field are
\begin{align}
m_{h^{\prime}}^2&=2v_{1}^2(\lambda_{1}+\lambda_{4})\cos^2\theta+4v_{2}^2(\lambda_{2}+\lambda_{4})\sin^2\theta+4\sqrt{2}\cos\theta\sin\theta v_{1}v_{2}\lambda_{4}\;.\nonumber \\
m_{h_{+}^{\prime}}^2&=2v_{1}^2(\lambda_{1}+\lambda_{4})\sin^2\theta+4v_{2}^2(\lambda_{2}+\lambda_{4})\cos^2\theta-4\sqrt{2}\cos\theta\sin\theta v_{1}v_{2}\lambda_{4}\;.
\end{align}

The other physical scalars of $m_{H^{\prime}}$ and $m_{H_{+}^{\prime}}$  include the CP-odd and charged components, which are not pertinent to our current discussion.  As mentioned in the main text,
these components are sufficiently  heavy to neglect the corresponding physical effects.

For the triplet and singlet scalars, after symmetry breaking the last line in Eq.~(\ref{potential}) will become  
\begin{align}
\mathcal{L}_{m}&=\frac{s^2}{2}\left(\frac{1}{2}\mu_{s}^2+3\lambda_{s}v_{s}^2\right)+\tilde{M}_{\Delta_{i}}^2(\Delta_{i}^{-}\Delta_{i}^{+}+(\Delta_{i}^{0})^2+\Delta^{--}_{i}\Delta_{i}^{++})+\tilde{\lambda}_{2i}(\Delta_{i}^{-}\Delta_{i}^{+}+(\Delta_{i}^{0})^2+\Delta^{--}_{i}\Delta_{i}^{++})^2\nonumber \\
&+\tilde{\lambda}_{3i}\left[\left(\frac{\Delta_{i}^{-}\Delta_{i}^{+}}{2}+\Delta_{i}^{0}\right)^2+\left(\Delta_{i}^{--}\Delta_{i}^{++}+\frac{\Delta_{i}^{-}\Delta_{i}^{+}}{2}\right)^2+
(\Delta_{i}^{-}\Delta_{i}^{++}-\Delta_{i}^{0}\Delta_{i}^{+})(\Delta_{i}^{--}\Delta_{i}^{+}-\Delta_{i}^{-}\Delta_{i}^{0})\right]
\end{align}
It will contribute the relevant mass as
\begin{align}
m_{s}^{2}&=\frac{1}{2}\mu_{s}^2+3\lambda_{s}v_{s}^2\;,\nonumber \\
m_{\Delta_{i}^{0}}^{2}&=\tilde{M}_{\Delta_{i}}^2+3(\tilde\lambda_{2i}+\tilde\lambda_{3i})\tilde{v_{i}}^2\;,\nonumber \\
m_{\Delta_{i}^{\pm}}^{2}&=\tilde{M}_{\Delta_{i}}^2+(\lambda_{2i}+\lambda_{3i})\tilde{v_{i}}^2\;,\nonumber \\
m_{\Delta_{i}^{\pm\pm}}^{2}&=\tilde{M}_{\Delta_{i}}^2+\tilde{\lambda}_{2i}\tilde{v}_{i}^2
\end{align}
We found that the mass difference for $\Delta^0,\Delta^+,\Delta^{++}$ depends on $\tilde v$. Due to the stringent constraints $\tilde v<O$(GeV) from EW $\rho$ parameter, we can naturally regard the triplet components as degenerate, $m_{\Delta^0}\approx m_{\Delta^+}\approx m_{\Delta^{++}}$.

\end{document}